%% file: cuMF.tex
\documentclass{vldb}
\usepackage{graphicx}
\usepackage{balance}  
\usepackage{multirow,tabularx, array, color,url, subfig}
\usepackage{algorithm}
\usepackage{algpseudocode}
\usepackage{hyperref}
\usepackage{tabularx}
\usepackage[utf8]{inputenc}
\usepackage{amsmath}
\usepackage{threeparttable}
\usepackage{listings}
\usepackage{color}
\usepackage{lipsum}
\usepackage{enumitem}
\setlist{itemsep=0pt}

\definecolor{codegreen}{rgb}{0,0.6,0}
\definecolor{codegray}{rgb}{0.5,0.5,0.5}
\definecolor{codepurple}{rgb}{0.58,0,0.82}
\definecolor{backcolour}{rgb}{0.95,0.95,0.92}
\lstdefinestyle{mystyle}{
    backgroundcolor=\color{backcolour},   
    commentstyle=\color{codegreen},
    keywordstyle=\color{magenta},
    numberstyle=\tiny\color{codegray},
    stringstyle=\color{codepurple},
    basicstyle=\footnotesize,
    breakatwhitespace=false,         
    breaklines=true,                 
    captionpos=b,                    
    keepspaces=true,                 
    numbers=left,                    
    numbersep=5pt,                  
    showspaces=false,                
    showstringspaces=false,
    showtabs=false,                  
    tabsize=2
}
 
\begin{document}

\algblock{ParFor}{EndParFor}
\algnewcommand\algorithmicparfor{\textbf{parfor}}
\algnewcommand\algorithmicpardo{\textbf{do}}
\algnewcommand\algorithmicendparfor{\textbf{end\ parfor}}
\algrenewtext{ParFor}[1]{\algorithmicparfor\ #1\ \algorithmicpardo}
\algrenewtext{EndParFor}{\algorithmicendparfor}


\title{Faster and Cheaper: Parallelizing Large-Scale Matrix Factorization on GPUs}



%
%
%
%

\numberofauthors{3} 

\author{
%
%
  \alignauthor
  Wei Tan\\
         \affaddr{IBM T. J. Watson Research Center}\\
         \affaddr{Yorktown Heights, NY, USA}\\
         \email{wtan@us.ibm.com}
  \alignauthor 
	Liangliang Cao\titlenote{Work done while the author worked at IBM.}\\
          \affaddr{Yahoo! Labs}\\
           \affaddr{New York City, NY, USA}\\
         \email{liangliang@yahoo-inc.com}
\alignauthor
 Liana Fong\\
          \affaddr{IBM T. J. Watson Research Center}\\
         \affaddr{Yorktown Heights, NY, USA}\\
         \email{llfong@us.ibm.com}
}

\maketitle

\begin{abstract}

Matrix factorization (MF) is employed by many popular algorithms, e.g., collaborative filtering. 
The emerging GPU technology, with massively multicore and high intra-chip memory bandwidth but limited memory capacity, presents an opportunity for accelerating MF much further when appropriately exploiting the GPU architectural characteristics.
This paper presents cuMF, a CUDA-based matrix factorization library
that implements  memory-optimized  alternate least square (ALS) method 
to solve very large-scale MF.  
CuMF uses a variety set of  techniques to maximize the performance on either single or multiple GPUs. These techniques include smart access of sparse data leveraging GPU memory hierarchy, using data parallelism in conjunction with model parallelism, 
minimizing the communication overhead between computing units, 
and utilizing a novel topology-aware parallel reduction scheme.
With only a single machine with four Nvidia GPU cards, cuMF can be 6-10 times as fast, and 33-100 times as cost-efficient, compared with the state-of-art distributed CPU solutions. Moreover, this cuMF can solve the largest matrix factorization problem ever reported yet in current literature, while maintaining impressively good performance.

\end{abstract}



\input{introduction}
\input{problem}
\input{memory}
\input{scaleup}
\input{exp}
\input{relatedwork}
\input{conclusion}
\balance

\bibliographystyle{abbrv}
\bibliography{cumf} 
\end{document}

%% file: introduction.tex
\section{Introduction}

\begin{figure}
\center{\includegraphics[width=\linewidth]
        {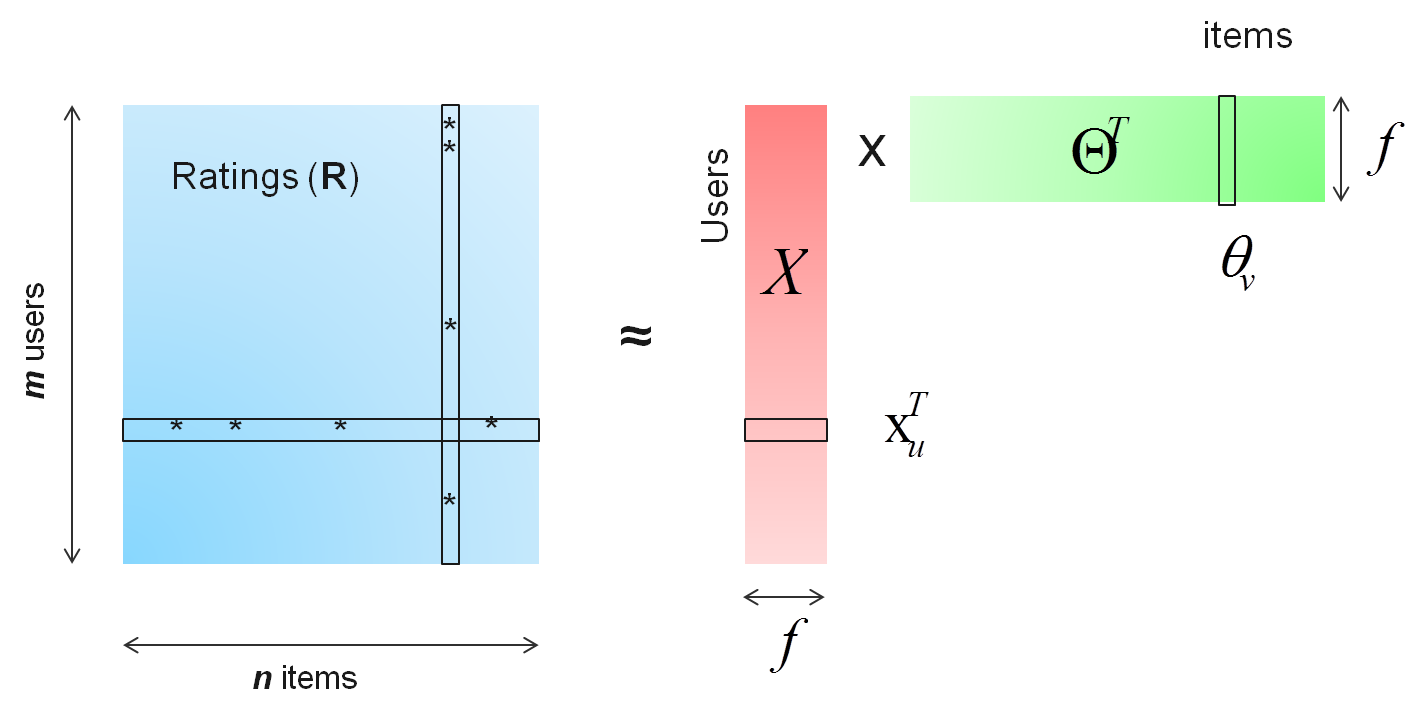}}
 \caption{Matrix factorization.}
\label{fig:mf}
\end{figure}

Sparse matrix factorization (SMF or MF) factors a sparse rating matrix $R$ ($m$ by $n$, with $N_z$ non-zero elements) into a $m$-by-$f$ and a $f$-by-$n$ matrices, as shown in Figure~\ref{fig:mf}.
MF is widely used for collaborative-filtering-based recommendations~\cite{mf-computer09} in e-commerce (e.g., Amazon) and digital content streaming (e.g., Netflix). Very recently, MF is also applied in text mining, deriving hidden features of words~\cite{pennington2014glove}. 

\begin{figure}
\center{\includegraphics[width=0.95\columnwidth]
        {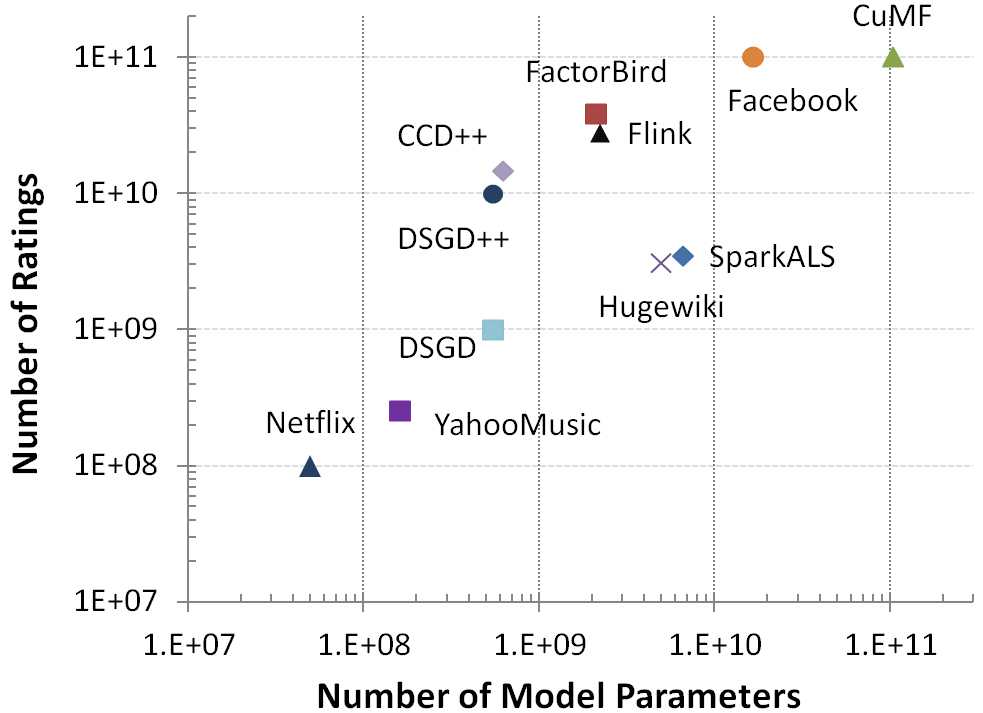}}
 \caption{The scale of MF data sets\protect\footnotemark. Y-axis is the $N_z$ of $R$, and x-axis is $(m+n)\times f$. CuMF can tackle MF problems of greater size, compared with existing systems.}
\label{fig:scale}
\end{figure}

Given the widespread use of MF, a scalable and speedy implementation is very important. In terms of \textbf{scale}, many parallel solutions~\cite{libmf-13,ccd++-icdm12,nomad14} target at medium-sized problems such as the Netflix challenge \cite{netflix08}. However, the industry-scale recommendation problems have evolved to two-orders-of-magnitude larger. Figure \ref{fig:scale} shows the scale of MF problems, in terms of number of ratings and number of model parameters. As an example, Facebook's MF is with $100+$ billion ratings, 1 billion users, and millions of items~\cite{facebook15}. No existing system except \cite{facebook15} has tackled problems at this scale. In terms of \textbf{speed}, recommendations need to evolve promptly in online applications. Current approaches including MPI \cite{nomad14}, Spark \cite{sparkals14} and parameter server \cite{factorbird14} address large-scale MF problems. However, they require costly clusters (e.g., 50-node) and still suffer from long latency.
\footnotetext{CCD++ \protect\cite{ccd++-icdm12}, DSGD \protect\cite{DSGD-kdd11}, DSGD++ \protect\cite{DBLP:conf/icdm/TeflioudiMG12}, Facebook \cite{facebook15}, Factorbird \protect\cite{factorbird14}, Flink \protect\cite{flink15}, Hugewiki \protect\cite{libmf-13}, Netflix \protect\cite{netflix08} SparkALS \protect\cite{sparkals14}, and YahooMusic \protect\cite{kdd-cup-yahoomusic-11}.}

Recently, the GPU emerges as an accelerator for parallel algorithms~\cite{GeMTC2014,hpdc2015gpu}. It has big compute power (typically 10x floating-point operations per second, flops vs. a CPU) and memory bandwidth (typically 5x vs. a CPU) \cite{hennessy2011computer}, but limited amount of control logic and memory capacity.
Particularly, GPU's success in deep learning \cite{costshpc2013} inspires us to try it for MF. In deep learning, the computation is mainly dense matrix multiplication which is \textbf{compute bound}. As a result, GPU can train deep neural network 10x as fast as CPU by saturating its flops. However, unlike deep learning, a MF problem involves sparse matrix manipulation which is usually \textbf{memory bound}. 
Given this, we want to explore a MF algorithm and a system that can still leverage GPU's compute and memory capability. 
We identified that, the alternating least square (ALS) algorithm \cite{mf-computer09} for MF is inherently parallel so as to exploit thousands of GPU cores. Moreover, compared with stochastic gradient descent (SGD), ALS has advantage when $R$ is made up of implicit ratings and therefore cannot be considered sparse \cite{mf-computer09}. 

Based on these observations, we design and implement \textbf{cuMF} (CUDA Matrix Factorization), a scalable ALS solution on one machine with one or more GPUs. 
CuMF achieves excellent scalability and performance by making \textbf{the following contributions}.

(1) On a single GPU, MF is inherently sparse and memory bound and thus difficult to utilize GPU's compute power.  We optimize memory access in ALS by various techniques including reducing discontiguous memory access, retaining hotspot variables in faster memory, and aggressively using registers. By this means cuMF gets closer to the roofline performance of a single GPU. 

(2) On multiple GPUs, we add data parallelism to ALS's inherent model parallelism. Data parallelism needs a faster reduction operation among GPUs, leading to (3).

(3) We also develop an innovative topology-aware, parallel reduction method to fully leverage the bandwidth between GPUs. By this means cuMF ensures that multiple GPUs are efficiently utilized simultaneously.


\begin{table}
\begin{threeparttable}
\centering
\caption{Speed and cost of cuMF on one machine with four GPUs, compared with three distributed CPU systems, on the cloud}
\label{tbl:cumf}
\begin{tabular}{|p{0.075\textwidth}|p{0.08\textwidth}|p{0.055\textwidth}|p{0.065\textwidth}|p{0.045\textwidth}|p{0.045\textwidth}|}  \hline
\textbf{Baseline}& baseline config &\#nodes&price\break/node/hr& cuMF speed & cuMF cost\\\hline
NOMAD&m3.xlarge&32&\$0.27&\textbf{10x}&\textbf{3\%} \\ \hline
SparkALS&m3.2xlarge&50&\$0.53&\textbf{10x} &\textbf{1\%}\\ \hline
Factorbird&c3.2xlarge&50&\$0.42&\textbf{6x} & \textbf{2\%}\\ \hline
\end{tabular}
\begin{tablenotes}[para,flushleft]
Note: Experiment details are in Section~\ref{sec:exp}. NOMAD~\cite{nomad14} uses Hugewiki data and AWS servers; it used m1.xlarge which is now superseded by m3.xlarge by Amazon. Factorbird's node is similar to AWS c3.2xlarge \cite{factorbird14}.

  \end{tablenotes}
\end{threeparttable}
\end{table}

The resulting CuMF is competitive in both speed and monetary cost. Table~\ref{tbl:cumf} shows cuMF's speed and cost compared with three CPU systems, NOMAD (with Hugewiki data)~\cite{nomad14}, Spark ALS~\cite{sparkals14}, and Factorbird~\cite{factorbird14}. NOMAD and Spark ALS use Amazon AWS, and we pick an AWS node type similar to what Factorbird uses. CPU and GPU systems' cost is calculated by 
$(price~per~node~per~hr)*(\#nodes)*(execution~time)$, 
with unit price taken when submitting this paper\footnote{AWS price: https://aws.amazon.com/ec2/pricing/; GPU machine price: http://www.softlayer.com/gpu}. CuMF runs on one machine with two Nvidia K80 (four GPUs devices in total) from IBM Softlayer, with an amortized hourly cost of \$2.44. With faster speed and fewer machine requirement, cuMF's overall cost of running these benchmarks is merely $1\%$-$3\%$ of the baseline systems compared. That is, cuMF is 33-100x as cost-efficient. 

In summary, this paper describes a novel implementation of MF on a machine with GPUs and the set of exemplary optimization techniques in leveraging the GPU architectural characteristics. The
experimental results demonstrate that with up to four Nvidia GPUs on one machine, cuMF is (1) competitive compared with multi-core methods, on medium-sized problems; (2) much faster than vanilla GPU implementations without memory optimization; (3) 6-10 times as fast, and 33-100 times as cost-efficient as distributed CPU systems, on large-scale problems; (4) more significantly, able to solve the largest matrix factorization problem ever reported. 

This paper is organized as follows. Section 2 formulates the problem of matrix factorization and explains the two challenges in large-scale ALS, i.e., memory access on one GPU and scalability on multiple GPUs. Section 3 introduces the memory-optimized ALS algorithm on a single GPU, to address challenge 1. Section 4 introduces the scale-up ALS algorithm to parallelize MF on multiple GPUs, to address challenge 2. Section 5 shows the experiment results and Section 6 reviews related work. Section 7 concludes the paper.

%% file: problem.tex
\section{Problem Definition}
\label{sec:background}

\subsection{ALS algorithm for matrix factorization}
\begin{table}
\begin{threeparttable}
\centering
\caption{Notations}
\label{tbl:notations}
\begin{tabular}{|c|p{5cm}|c|} \hline
\textbf{Name}	& \textbf{Meaning} & \textbf{Range}	\\ \hline
$R$ & sparse rating matrix: $m$ by $n$ &  \\ \hline
$X$ & low rank matrix: $m$ by $f$ &  \\ \hline
$\Theta$ & low rank matrix: $n$ by $f$ &  \\ \hline
$m$ & vertical dimension of $R$ & $10^3$ to $10^9$\\ \hline
$n$ & horizontal dimension of $R$ & $10^3$ to $10^9$ \\ \hline
$f$ & dimension of latent features & 5 to 100s \\ \hline
$N_z$ & number of non-zero entries in $R$ & $10^8$ to $10^{11}$ \\ \hline
$r_{uv}$ & $R$'s value at position $(u,v); 1\leq u\leq m, 1\leq v\leq n $ &  \\ \hline
$\textbf{x}_u^T$ & $X$'s $u$th row; $1\leq u\leq m$ &  \\ \hline
$\theta_v$ & $\Theta^T$'s $v$th column; $1\leq v\leq n$ &  \\ \hline
$R_{u*}$ & $R$'s $u$th row; $1\leq u\leq m$ &  \\ \hline
$R_{*v}$ & $R$'s $v$th column; $1\leq v\leq n$ &  \\ \hline
\end{tabular}
\begin{tablenotes}[para,flushleft]
Note: usually $N_z \gg m, n$ and $m, n \gg f$.
  \end{tablenotes}
\end{threeparttable}
\end{table}
Referring to the notations listed in Table~\ref{tbl:notations}, matrix factorization is to factor a sparse matrix $R$ with two lower-rank, dense matrices $X$ and $\Theta$, such that:
\begin{center}
$R\approx X \cdot \Theta^{T}$
\end{center}

As illustrated in Figure \ref{fig:mf}, suppose $r_{uv}$ is the non-zero element of $R$ at position $(u,v)$, we want to minimize the following cost function \eqref{eq-mf}. To avoid overfitting we use weighted-$\lambda$-regularization proposed in \cite{netflix08}, where $n_{x_u}$ and $n_{\theta_v}$ denote the number of total ratings on user $u$ and item $v$, respectively.

\begin{equation}
 J = \sum_{u,v} (r_{uv} - \mathbf{x}_u^T\theta_v)^2
 +\lambda (\sum_{u}n_{x_u}||\mathbf{x}_u||^2 +\sum_{v}n_{\theta_v}||\mathbf{\theta}_v||^2)
\label{eq-mf}
\end{equation}

Many optimization methods, including ALS \cite{netflix08}, CGD \cite{ccd++-icdm12}, and SGD \cite{libmf-13} have been applied to minimize $J$. We adopt the ALS approach that would first optimize $X$ while fixing $\Theta$, and then to optimize $\Theta$ while fixing $X$. Consider 
\begin{align*}
{\frac{\partial J}{\partial \textbf{x}_u}=0} 
\end{align*}
and
\begin{align*}
{\frac{\partial J}{\partial \theta_v}=0} 
\end{align*}
which lead to the following equation:
\begin{align}
\label{eq:update-x}
 \sum\limits_{r_{uv}\neq0}
 (\theta_v \theta_v^T+\lambda I) \cdot \textbf{x}_u  &= \Theta^T\cdot R_{u*}^T
\end{align}
together with:
\begin{align}
\label{eq:update-theta}
 \sum\limits_{r_{uv}\neq0}
 (\textbf{x}_u \textbf{x}_u^T+\lambda I) \cdot \theta_v &= X^T\cdot R_{*v}
\end{align}


By this means, ALS updates $X$ using eq.~\eqref{eq:update-x}, and updates $\Theta$ using eq.~\eqref{eq:update-theta}, in an alternating manner. Empirically, ALS often converges in 5-20 iterations, with each iteration consisting of both update-$X$ and update-$\Theta$. In the rest of this paper, we explain our method using update-$X$. The same method is applicable to update-$\Theta$.

The formalism of ALS enables solving in parallel so as to harness the power of GPU. 
Eqs.~\eqref{eq:update-x} and~\eqref{eq:update-theta} shows that, the updates of each $\textbf{x}_u$  and $\theta_v$ are independent of each other. This independent nature does not hold for SGD, which randomly selects a sample $r_{uv}$, and updates the parameters by:
\begin{align}\label{eq-sgd}
\textbf{x}_u & = \textbf{x}_u - \alpha [ (\textbf{x}_u^T \theta_v - r_{uv}) \theta_v + \lambda \textbf{x}_u] \nonumber\\
\theta_v & = \theta_v - \alpha [ (\textbf{x}_u^T \theta_v - r_{uv})\textbf{x}_u + \lambda \theta_v]
\end{align}

Suppose there are two random samples $r_{uv}$ and $r_{uv'}$ with the same row index $u$, their updates to $\textbf{x}_u$ cannot be treated independently. Previous works on CPUs ~\cite{libmf-13, nomad14, DSGD-kdd11, DBLP:conf/icdm/TeflioudiMG12} all partition $R$ into blocks with no overlapping rows and columns. Such a strategy works effectively on tens of CPU cores but is difficult to scale to a GPU with thousands of cores. 
As a result, we choose ALS instead of SGD for cuMF.

\subsection{Challenges of speedy and scalable ALS}
\label{sec:als-challenges}
\begin{table*}
\begin{threeparttable}
\centering
\caption{Compute cost and memory footprint of ALS: the update-$X$ step}
\label{tbl:cost}
\begin{tabular}{ |l|l|l|l|l|l| } \hline 
& & \multicolumn{2}{c}{\textbf{compute cost}} & \multicolumn{2}{|c|}{\textbf{memory footprint}} \\ \hline 
& & $A_u$ in~\eqref{eq:update-x} & $B_u$ in~\eqref{eq:update-x}& $A_u$ in~\eqref{eq:update-x}& $B_u$ in~\eqref{eq:update-x}\\ \hline 
\multirow{3}{*}{\textbf{get\_hermitian\_x}}& one item & $N_zf(f+1)/2m$  & $(N_z+N_zf)/m+2f$ & $f^2$ & $nf+f+(2N_z+m+1)/m$\\
& $m_b$ items & $m_bN_zf(f+1)/2m$  & $m_b(N_z+N_zf)/m+2m_bf$ & $m_bf^2$ & $nf+m_bf+m_b(2N_z+m+1)/m$\\
& all $m$ items & $N_zf(f+1)/2$  & $N_z+N_zf+2mf$ & $mf^2$ & $nf+mf+(2N_z+m+1)$\\
 \hline 
 \hline
\multirow{3}{*}{\textbf{batch\_solve}}& one item &$f^3$&&&\\
& $m_b$ items & $m_bf^3$ & &&\\
& all $m$ items & $mf^3$ & &&\\
\hline
\end{tabular}
\begin{tablenotes}[para,flushleft]
Note: here we omit some minor computations and auxiliary data structures needed in eq.~\eqref{eq:update-x}.
  \end{tablenotes}
\end{threeparttable}
\end{table*}

Table~\ref{tbl:cost} lists the compute cost and memory footprint of solving $X$ with eq.~\eqref{eq:update-x}, using single precision. The calculation is divided into two phases, i.e., 

\textbf{get\_hermitian\_x} to obtain the left-hand Hermitian matrix $A_u=\sum\limits_{r_{uv}\neq0}(\theta_v \theta_v^T+\lambda I)$ and the right-hand $B_u=\Theta^T\cdot R_{u*}^T$, and

\textbf{batch\_solve} to solve many equations $A_u\textbf{x}_u=B_u$. 

In line 3 of Table~\ref{tbl:cost}: \textit{one item} in phase get\_hermitian\_x, to solve one row $\textbf{x}_u$, obtaining $A_u$ needs to calculate $N_z/m$ times\footnote{$N_z/m$ is the average number of non-zero entries per row.} of $\theta_v \theta_v^T$s, each of which needs $f(f+1)/2$ multiplications. The cost of obtaining $B_u$ is $(N_z+N_zf)/m+2f$ \cite{cusparse}. 
In terms of memory, $A_u$ uses $f^2$ floats, $B_u$ uses $f$, $\Theta^T$ uses $nf$, and a row of $R$ in Compressed Sparse Row (CSR) format uses $(2N_z+m+1)/m$. In phase batch\_solve, solving the linear equation $A_u\textbf{x}_u=B_u$ does not need additional memory storage by using in-place solvers, but has an $f^3$ computation cost. 

\noindent\textbf{Challenge 1. On a single GPU, how to optimize sparse, irregular and intensive memory access.}

Table~\ref{tbl:cost} shows that, computation is bounded in both phases \textbf{get\_hermitian\_x} ($\mathcal{O}(N_zf^2)$) and \textbf{batch\_solve} ($\mathcal{O}(mf^3)$). CUDA library cuBLAS~\cite{cublas} already provides dense solvers for phase batch\_solve, so we focus on the get\_hermitian\_x phase. This phase is very costly, especially when $N_z\gg m$ and therefore $N_zf^2>mf^3$. What is more troublesome is the \textit{sparse}, \textit{irregular} and \textit{intensive} memory access in this phase. Details are as follows:
\begin{enumerate}
\item Access many columns $\theta_v$ subject to $r_{uv}\neq 0$ for every $u$. This access is \textit{irregular} w.r.t. $\Theta^T$, due to the sparseness of $R$. In each iteration to solve one $\textbf{x}_u$ we need to access $n_{x_u}$ columns ($N_z/m$, on average) spread \textbf{sparsely} and \textbf{discontiguously} across the $n$ columns of $\Theta^T$. For example, in the Netflix data set \cite{netflix08}, one user rates around 200 items on average, leading to a discontiguous access of 200 columns from the total 17,770 in $\Theta^T$.
\item Aggregate many $\theta_v \theta_v^T$s and $\textbf{x}_u \textbf{x}_u^T$s, is memory \textit{intensive} due to the large number of $\theta_v$s and $\textbf{x}_u$s to aggregate. According to eq. \eqref{eq:update-x}, obtaining $A_u$ needs to calculate many $\theta_v \theta_v^T$s and aggregate them. Therefore, each element in column vector $\theta_v$ is accessed frequently, and the partial aggregation result is updated frequently. To calculate $\theta_v\theta_v^T$ we need to read each element of $\theta_v$ $f$ times; after obtaining a $\theta_v\theta_v^T$, to add it to $\sum\limits_{r_{uv}\neq0}(\theta_v \theta_v^T+\lambda I)$ we need to write $f(f+1)/2$, or $f^2$ elements if the downstream solver does not appreciate symmetricity.
\end{enumerate}

Section \ref{sec:singlegpu} presents how cuMF tackles Challenge 1, with experiment results shown in Sections \ref{sec:exp-one-gpu} and \ref{sec:exp-mo}.

\noindent\textbf{Challenge 2. On multiple GPUs, how to scale and minimize communication overhead.}

When $m$, $n$, $N_z$ and $f$ get larger, ALS is bounded by the memory capacity of a single GPU. For example, the update-$X$ iteration is to be bounded by memory footprint of $m$ $A_u$s ($mf^2$ without considering symmetricity), $X^T$ ($mf$), $\Theta^T$ ($nf$) and $R$ ($2N_z+m+1$). The current Nvidia Maxwell and Kepler GPUs have 12 GB memory per device. Each device would only be able to load 3 billion ($3\times 10^9$) single precision floats. However, the smallest data set, i.e., Netflix, in Figure \ref{fig:scale}, has $m=480$K. When $f=100$, $m$ Hermitian matrices are with size $mf^2=480$K$\times 100^2=4.8$ billion floats $>$ 3 billion.

Previous CPU solutions already encountered and partially addressed this memory capacity issue. PALS~\cite{netflix08} partitions $X$ and $R$ by rows, solving each partition in parallel by replicating $\Theta^T$. However, this \textbf{model parallelism} is only feasible when $\Theta^T$ is small. SparkALS~\cite{sparkals14}, the ALS implementation in Spark MLlib~\cite{sparkmllib15}, also partitions $X$ and $R$ by rows, and then solve each partition $X_i$ in parallel. Its improvement to PALS is that, instead of replicating $\Theta^T$, it splits $\Theta^T$ into overlapping partitions $\{\Theta^T_i\}$, where $\Theta^T_i$ contains only the necessary $\theta_v$ columns for all $\textbf{x}_u$s in $X_i$. This improvement still has several deficiencies:
\begin{enumerate}
	\item Generating ${\Theta^T_i}$ from ${X_i}$ is actually a graph partitioning task and time consuming.
	\item Transferring each $\Theta^T_i$ to $X_i$ involves much network traffic, especially when $N_z \gg m$.
	\item $\Theta^T_i$ may still be too big to fit into a single GPU device, especially when $N_z \gg m$.
\end{enumerate}

Section \ref{sec:mgpu} presents how cuMF tackles Challenge 2, with experiment results shown in Sections \ref{sec:exp-scalability} and \ref{sec:exp-xlarge}.


%% file: memory.tex
\section{Memory-optimized ALS on one GPU}
\label{sec:singlegpu}
\subsection{The GPU memory hierarchy}
To address \textbf{Challenge 1} ``On a single GPU, how to optimize sparse, irregular and intensive memory access'', we need direct control on GPU's memory hierarchy. We choose Nvidia GPUs because they provides a rich set of \textit{programmable memory} of different characteristics, shown in Table \ref{tbl:gpu-mem}. \footnote{There are \textit{non-programmable memory} such as L1 and L2 cache. They also accelerate memory access but are not directly controllable by programmers. Therefore in cuMF we focus on the optimization by using the programmable memory.}

\begin{table}[h!]
\centering
\caption{Programmable GPU memory}\label{tbl:gpu-mem}
\begin{tabular}{|c|p{1cm}|p{1.2cm}|p{2.6cm}|} \hline
\textbf{Memory type}	& \textbf{Size} & \textbf{Latency} & \textbf{Scope} \\ \hline
\textit{global} & large & high & application\\ \hline
\textit{texture} & medium &  medium & application, read-only \\ \hline
\textit{shared} & small 	& low & thread block \\ \hline
\textit{register} &	small	& lowest & thread; not indexable \\ \hline
\end{tabular}
\end{table}

Although the principles of memory optimization are generally known, the specific implementation of ALS on GPU is not trivial due to the following reasons:
\begin{enumerate}
  \item GPU has a lower clock frequency than CPU (typically $<1$ GHz vs. 2-3 GHz). If the massive parallelism in GPU is not fully utilized, cuMF is likely to be slower than the highly-optimized CPU implementations.
  \item Compared with CPU, GPU's global memory is smaller, e.g., 12 GB. In contrast, GPU has a much larger register file, e.g., 4 MB, which is largely ignored nowadays.
  \item The control of register, shared, texture and global memory is complex. The global memory is large but slow, texture memory is read-only, and register and shared memory are not visible across GPU kernels (i.e., device functions). Moreover, registers are not \textit{dynamically indexable}, which prevents them from being used for large arrays.  
\end{enumerate}

Due to these difficulties, without insight on both GPU hardware and algorithm specifics, an implementation can easily be bounded by memory capacity, latency or bandwidth, preventing us from harnessing the full power of GPU.

\subsection{The base ALS algorithm}
The base ALS algorithm \ref{alg:base-als} shows how to update $X$ with eq. \eqref{eq:update-x}. The algorithm to update $\Theta$ is similar with all variables symmetrically exchanged. Algorithm \ref{alg:base-als} consists of two procedures: \textsc{Get\_Hermitian\_X()} and \textsc{Batch\_Solve()}.

\begin{algorithm}[h!]
\caption{Base ALS: Update $X$
\newline \textbf{Input} $R_{m \times n}$
\newline \textbf{Input} $\Theta^T: [\theta_1, \theta_2, ..., \theta_n]_{f \times n}$
\newline \textbf{Output} $X: [\textbf{x}_1^T; \textbf{x}_2^T; ...; \textbf{x}_m^T]_{m \times f}$
}
\label{alg:base-als}
\begin{algorithmic}[1]
\Procedure{Get\_Hermitian\_X}{$R,\Theta^T$}
\For{$u \gets 1, m$}
\State $\Theta^T_u \gets$ sub-matrix of $\Theta^T$ with cols $\theta_v$ s.t. $r_{uv}\neq 0$\label{line:3}
\State $A_u \gets 0$
\ForAll {columns $\theta_v$ in  $\Theta^T_u$}
    \State $A_u \gets A_u+\theta_v\theta^T_v+\lambda I$ \label{line:6}
\EndFor
\State $B_u \gets \Theta^{T} \cdot R_{u*}^T$
\EndFor
\State \textbf{return} $([A_1,A_2,...A_m], [B_1,B_2,...,B_m])$
\EndProcedure

\Statex

\Procedure{Batch\_Solve}{$[A_1,A_2,...A_m], [B_1,B_2,...,B_m]$}
\For{$u \gets 1, m$}
\State $\textbf{x}_u \gets$ solve $A_{u}\cdot \textbf{x}_u=B_u$
\EndFor
\State \textbf{return} $[\textbf{x}_1,\textbf{x}_2,...\textbf{x}_m]^T$
\EndProcedure

\Statex

\State $(A, B) \gets$ \Call{Get\_Hermitian\_X}{$R, \Theta^T$}
\State $X \gets$ \Call{Batch\_Solve}{$A, B$} 
\end{algorithmic}
\end{algorithm}

\subsection{The memory-optimized ALS algorithm MO-ALS}
Table~\ref{tbl:cost} indicates that, \textsc{Get\_Hermitian\_X()} in Algorithm \ref{alg:base-als} is memory intensive. We observed that Lines 3-7, i.e., computing $A_u$, takes much of the overall execution time. To optimize the performance, we enhance Algorithm \ref{alg:base-als} by leveraging different types of GPU memory.  We call this memory-optimized ALS algorithm \textbf{MO-ALS}, as described in Algorithm~\ref{alg:mo-als}.  The following lines in Algorithm \ref{alg:base-als} are enhanced in MO-ALS: 
\begin{enumerate}
\item{Reading from $\Theta^T$ in Line \ref{line:3}}. $\Theta^T$ with dimension $f \times n$ is stored in global memory. When collecting the sub-matrix $\Theta_u^T$ from $\Theta^T$, we use \textbf{texture memory} as the cache because: (1) this collecting process enjoys spatial locality, (2) $\Theta^T$ is read-only when updating $X$, and (3) different $\Theta_u^T$s can potentially re-use the same $\theta_v$s cached in texture memory. As a result, this caching step reduces discontiguous memory access. This optimization is shown in Line 3 in Algorithm ~\ref{alg:mo-als}.

\item{Storage of $\Theta^T_u$ in Line \ref{line:3}}. We use one thread block with $f$ threads to calculate each $A_u$, and use the per-block \textbf{shared memory} to store $\Theta_u^T$, so as to speed up the subsequent read in Line \ref{line:6}. However, for each $A_u$, we are not able to copy the whole $\Theta_u^T$ into its shared memory space. This is because $\Theta_u^T$ is of size $f \times n_{x_u}$ and too big compared to the 48 or 96 KB per-SM\footnote{SM or SMX: stream multiprocessor. A GPU device usually consists of 10 to 15 SMs.} shared memory. If a single thread block consumes too much shared memory, other blocks are prohibited from launching, resulting in low parallelism. In order to achieve higher parallelism, we select a bin size $bin$, and for each ${\textbf{x}_u}$ only allocate a share memory space $\Theta_u^T[bin]$ of size $f \times bin$. In practice we choose $bin$ between 10 and 30, while $n_{x_u}$ can be hundreds to thousands. We iteratively move a subset of $\Theta_u^T$ into $\Theta_u^T[bin]$ to be processed in the following step. This optimization is shown in Lines 5-10 in Algorithm ~\ref{alg:mo-als}.

\item{Update of $A_u$ in Line \ref{line:6}}. Here we need to read a $\theta_v$ from $\Theta_u^T[bin]$, calculate the $f \times f$ elements of $\theta_v \cdot \theta^T_v$, and add them to global variable $A_u$. Obviously $A_u$ is a memory hotspot. In order to speedup the aggregation in $A_u$, we choose \textbf{register memory} to hold $\sum\limits_{\theta_v \in \Theta_u^T[bin]}\theta_v \theta_v^T$, and only update global memory $A_u$ after we iterate over all columns in $\Theta_u^T$. This reduces global memory access by a factor of $n_{x_u}$. This optimization is shown in Line 8 in Algorithm ~\ref{alg:mo-als}. More details are discussed in the following Section \ref{sec:register}.
\end{enumerate}

\begin{figure*}[t]
\center{\includegraphics[width=0.65\linewidth]
        {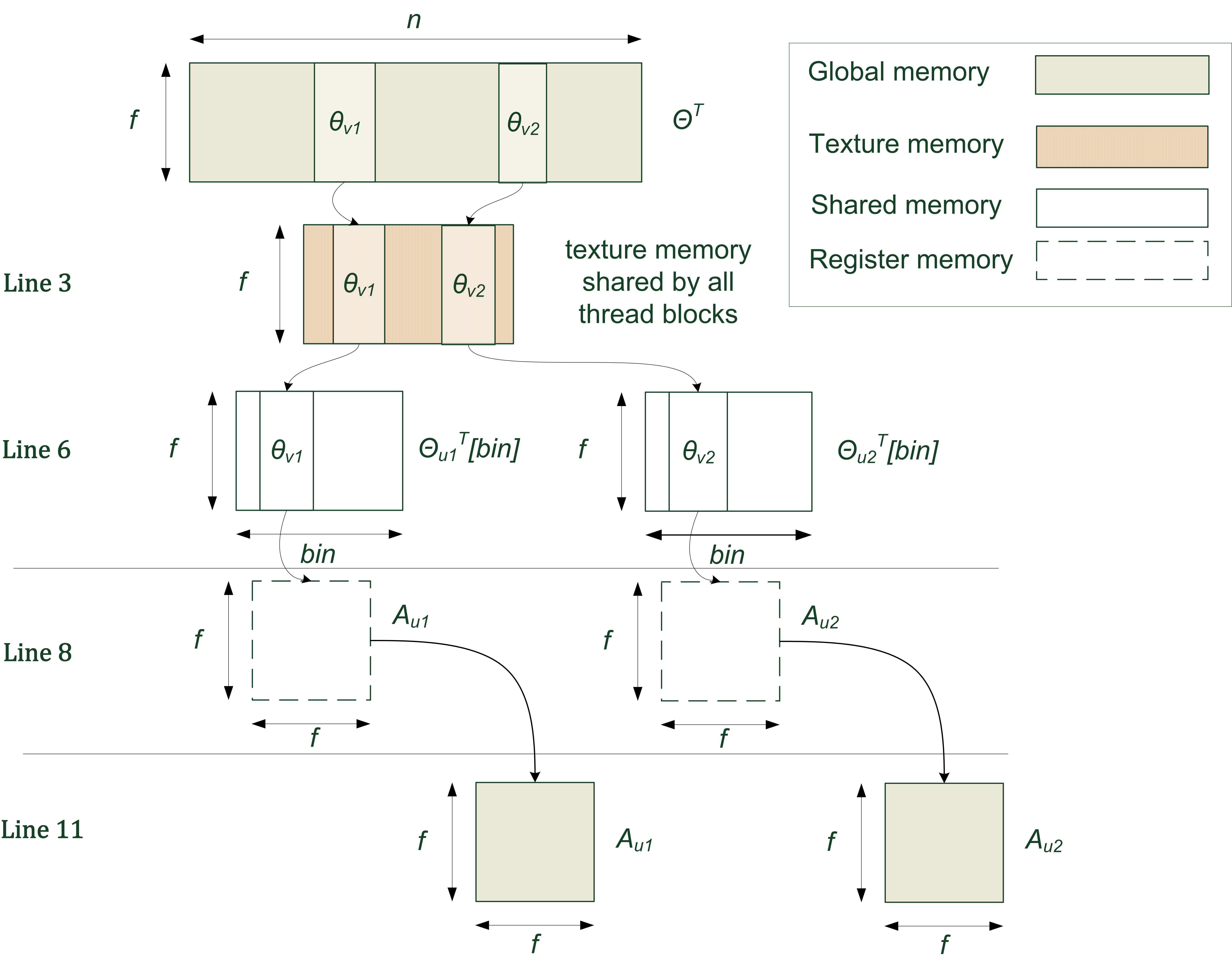}}
 \caption{Illustration of memory usage in MO-ALS. Line numbers correspond to those in Algorithm \ref{alg:mo-als}. For simplicity, we solve two rows of $X$, i.e., $\textbf{x}_{u1}$ and $\textbf{x}_{u2}$, in parallel. In reality we solve as many rows of $X$ as possible in parallel.}
\label{fig:gpu-memory}
\end{figure*}

Figure~\ref{fig:gpu-memory} illustrates the memory usage of MO-ALS.

\begin{algorithm}[h!]
\caption{MO-ALS: Memory-Optimized ALS; update $X$ on one GPU.
\newline $\mathcal{G}\{ var \}$: $var$ in global memory 
\newline $\mathcal{T}\{ var \}$: $var$ in texture memory
\newline $\mathcal{S}\{ var \}$: $var$ in shared memory
\newline $\mathcal{R}\{ var \}$: $var$ in register memory
\newline \textbf{Input} $R_{m \times n}$
\newline \textbf{Input} $\Theta^T: [\theta_1, \theta_2, ..., \theta_n]_{f \times n}$
\newline \textbf{Output} $X: [\textbf{x}_1^T; \textbf{x}_2^T; ...; \textbf{x}_m^T]_{m \times f}$
}
\label{alg:mo-als}
\begin{algorithmic}[1]
\Procedure{Get\_Hermitian\_X\_MO}{$R,\Theta^T$}
\For{$u \gets 1, m$}
\State $\mathcal{T}\{\Theta^T_u\} \gets$ sub-matrix of $\mathcal{G}\{\Theta^T\}$ with cols $\theta_v$ s.t. $r_{uv}\neq 0$ \label{line:2-1}
\State $\mathcal{R}\{A_u\} \gets 0$
\While {$\mathcal{T}\{\Theta^T_u\}$ has more cols not processed}
\State $\mathcal{S}\{\Theta_u^T[bin]\} \gets$ next $bin$ cols from $\mathcal{T}\{\Theta^T_u\}$
\ForAll {cols $\theta_v$ in $\mathcal{S}\{\Theta^T_u[bin]\}$}
    \State $\mathcal{R}\{A_u\} \gets \mathcal{R}\{A_u\}+\mathcal{S}\{\theta_v\}\mathcal{S}\{\theta^T_v\}+\lambda I$ \label{line:2-2}
\EndFor
\EndWhile
\State $\mathcal{G}\{A_u\} \gets \mathcal{R}\{A_u\}$
\State $\mathcal{G}\{B_u\} \gets \mathcal{G}\{\Theta^{T}\} \cdot \mathcal{G}\{R_{u*}^T\}$
\EndFor
\State \textbf{return} $\mathcal{G}([A_1,A_2,...A_m], [B_1,B_2,...,B_m])$
\Statex
\State $(A, B) \gets$ \Call{Get\_Hermitian\_X\_MO}{$R, \Theta^T$}
\State $X \gets$ \Call{Batch\_Solve}{$A, B$} 
\EndProcedure
\end{algorithmic}
\end{algorithm}

\subsection{Enhanced utilization of registers}
\label{sec:register}
We exploit the GPU register file which is larger and has higher bandwidth compared to its shared memory \cite{DBLP:conf/emnlp/CannyHK13}. For example, in the latest Nvidia Maxwell generation GPUs, each SM has a 256 KB register file and only 96 KB shared memory. However, while there is much focus on using shared memory \cite{optimizecuda08}, the use of registers is surprisingly ignored. This \textbf{under-utilization of registers} is mainly due to the fact that, register variables cannot be dynamically indexed. That is to say, you cannot declare and refer to an array in register file\footnote{An exception is that, the CUDA compiler may put very small ($\leq 5$) arrays on registers in loop unfolding.}. In Algorithm \ref{alg:mo-als}, $A_u$ is with size $f^2$ and to put it in register and access it, we have to declare $f^2$ variables instead of a single array. This makes the CUDA code hard to write. We use macro expansion in C to generate such a verbose paragraph of code. The snippet in Listing \ref{register-code} demonstrates how the expanded code looks like when $f=10$. 
\begin{lstlisting}[label=register-code, caption=CUDA kernel code to use registers when \textit{f} is 10, language=C]
get_Au_kernel()
{ ...
  //declare Au in registers
  float temp0 = 0, temp1 = 0, temp2 = 0, 
  temp3 = 0, temp4 = 0, temp5 = 0, 
  temp6=0, temp7=0, temp8=0, temp9=0;
  ...
  float temp90 = 0, temp91 = 0, temp92 = 0, 
  temp93 = 0, temp94 = 0, temp95 = 0, 
  temp96=0, temp97=0, temp98=0, temp99=0;
  //aggregate Au in register 
  for(k){
  	temp0 += theta[k*f]*theta[k*f];
    temp1 += theta[k*f]*theta[k*f+1];
    ...
    temp98 += theta[k*f+9]*theta[k*f+8];
    temp99 += theta[k*f+9]*theta[k*f+9];
  }
  //copy register to global memory
  Au[0] = temp0;
  Au[1] = temp1;
  ...
  Au[98] = temp98;
  Au[99] = temp99;
}
\end{lstlisting}

\textbf{Limitation of MO-ALS.} Algorithm \ref{alg:mo-als} is able to deal with big $X$ with one GPU, as long as $\Theta$ can fit into it. When $X$ is big and $\Theta$ is small, we first load the whole $\Theta$ to the GPU, then load $R$ and solve $X$ in batches. However, this batch-based approach does not work when $\Theta$ cannot fit into a single GPU. This motivates us to scale to multiple GPUs on a single machine, as presented in Section \ref{sec:mgpu}.  

%% file: scaleup.tex
\section{Scale-up ALS on multiple GPUs}
\label{sec:mgpu}
Section~\ref{sec:singlegpu} addresses \textbf{Challenge 1} regarding memory optimization on a single GPU. As problem size gets bigger, we need to address \textbf{Challenge 2}: ``On multiple GPUs, how to scale and minimize communication overhead.''
This section presents a scale-up algorithm called \textbf{SU-ALS} which adds \textbf{data-parallelism} and \textbf{parallel-reduction} on top of MO-ALS.

\subsection{The SU-ALS algorithm}
In distributed machine learning, \textbf{model parallelism} and \textbf{data parallelism} are two common schemes~\cite{jeffdean-dnn-nips12}. Model parallelism partitions \textbf{parameters} among multiple learners with each one learns a subset of parameters. Data parallelism partitions the training \textbf{data} among multiple learners with each one learns all parameters from its partial observation. These two schemes can be combined when both model parameters and training data are large. 

ALS is inherently suitable for model parallelism, as the updates of each $\textbf{x}_u$ and $\theta_v$ are independent. As discussed in Section~\ref{sec:als-challenges}, both PALS and SparkALS employ only model parallelism without considering data parallelism. To solve $X$ in parallel, PALS and SparkALS partition $X$ among multiple nodes. PALS broadcasts the whole $\Theta^T$ while SparkALS transfers a subset of it to each $X$ partition. As pointed out by \cite{ccd++-icdm12}, both approaches are inefficient and may cause out-of-memory failure, when $\Theta^T$ is big and ratings are skewed.

To tackle large-scale problems, on top of the existing model parallelism, we design a data-parallel approach. A limitation of model parallelism is that, it requires all $A_u$s in one partition $X^{(j)}$ ($1\leq j\leq q$) to be computed on the same GPU. Consequently, a subset of $\Theta^T$ has to be transferred into that GPU. In contrast, our data-parallel approach distributes the computation of any single Hermitian matrix $A_u$ to multiple GPUs. Instead of transferring all $\theta_v$s to one GPU, it calculates a local $A_u$ on each GPU with only the local $\theta_v$s, and reduce (aka., aggregate) many local $A_u$s later. Assume that there are \textit{p} GPUs to parallelize on, we re-write eq. \eqref{eq:update-x} to its data-parallelism form as:
\begin{align}
 A_u=\sum\limits_{r_{uv}\neq0} (\theta_v \theta_v^T+\lambda I) =  
\sum\limits_{i=1}^p\sum\limits_{r_{uv}\neq0}^{GPU_i} (\theta_v \theta_v^T+\lambda I)  
\end{align}

\begin{figure}
\center{\includegraphics[width=\columnwidth]
        {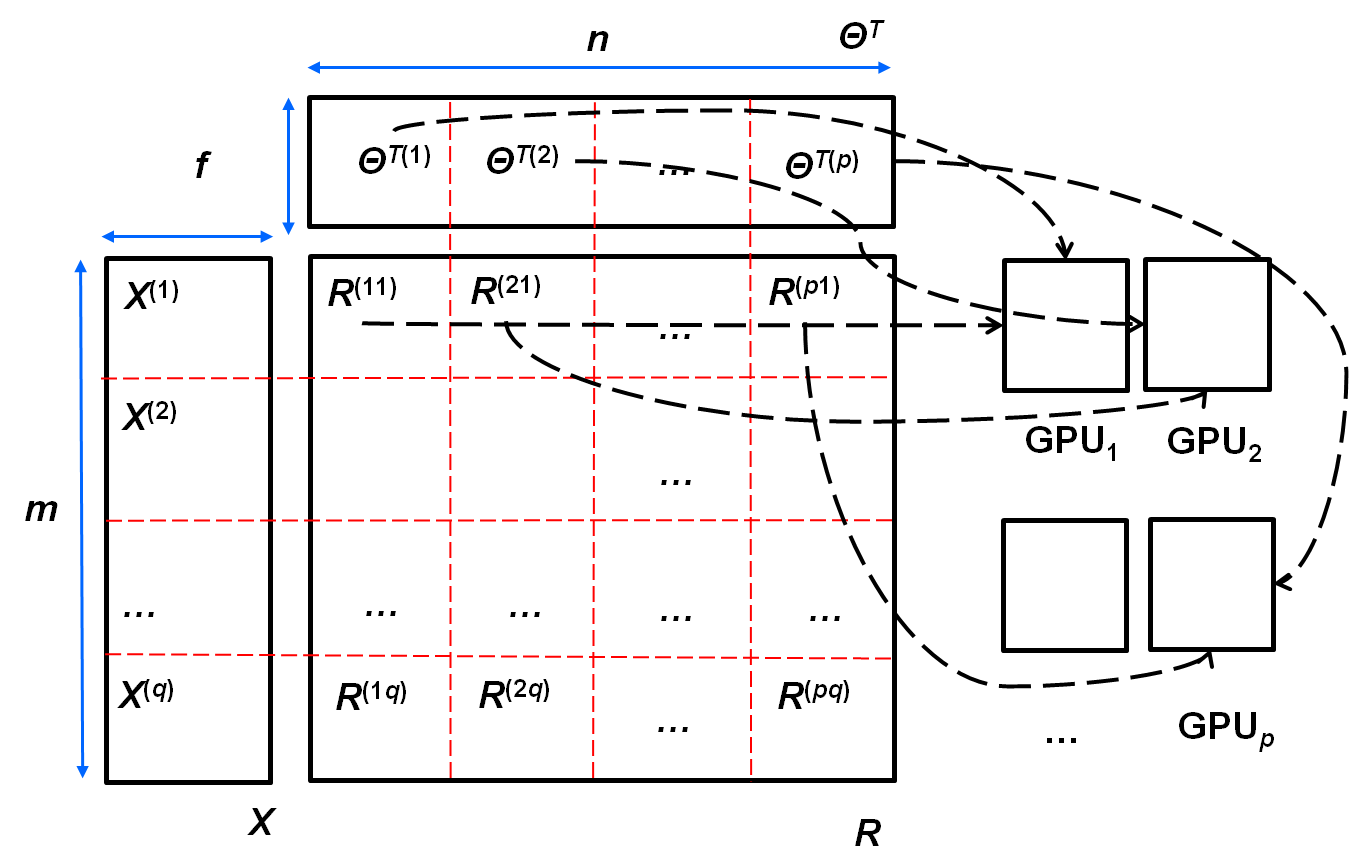}}
 \caption{SU-ALS. $\Theta^T$ is partitioned evenly and vertically, and stored on $p$ GPUs. $X$ is partitioned evenly and horizontally, and solved in batches, achieving model parallelism. Each $X$ batch is solved in parallel on $p$ GPUs, each with $\Theta^T$'s partition on it, achieving data parallelism.}
\label{fig:su-als}
\end{figure}

This approach is described in Algorithm~\ref{alg:su-als} and illustrated in Figure~\ref{fig:su-als}.

\begin{algorithm}[h!]
\caption{SU-ALS: Scale-Up ALS; update $X$ on multiple GPUs.
}
\label{alg:su-als}
\begin{algorithmic}[1]
\State Given $p$ GPUs: GPU$_1$, GPU$_2$, ..., GPU$_p$. 
\State $\{\Theta^{T(1)}, \Theta^{T(2)},...,\Theta^{T(p)}\} \gets VerticalPartition(\Theta^T, p)$
\State $\{X^{(1)}, X^{(2)},...,X^{(q)}\} \gets HorizontalPartition(X, q)$
\State $\{R^{(11)}, R^{(12)},...,R^{(pq)}\} \gets GridPartition(R, p, q)$
\ParFor{$i \gets 1, p$} \Comment{parallel copy to each GPU$_i$}
\State copy GPU$_i \gets \Theta^{T(i)}$
\EndParFor
\For{$j \gets 1, q$} \Comment{model parallel}
\ParFor{$i \gets 1, p$} \Comment{data parallel on GPU$_i$}
\State copy GPU$_i$$ \gets R^{(ij)}$ 
\State $(A^{(ij)}, B^{(ij)})\gets $ \Call{Get\_Hermitian\_X\_MO}{$R^{(ij)}, \Theta^{T(i)}$}
\State \Call{synchronize\_threads}{$ $}
\State $\{A^{(ij)}_1, A^{(ij)}_2, ..., A^{(ij)}_p\} \gets A^{(ij)}$
\State $\{B^{(ij)}_1, B^{(ij)}_2, ..., B^{(ij)}_p\} \gets B^{(ij)}$
\State $A^{(j)}_i \gets \sum\limits_{k=1}^p A^{(kj)}_i$
\State $B^{(j)}_i \gets \sum\limits_{k=1}^p B^{(kj)}_i$ 
\State $X^{(j)}_i \gets$ \Call{Batch\_Solve}{$A^{(j)}_i, B^{(j)}_i$}  
\EndParFor
\State $X^{(j)} \gets \{X^{(j)}_1,X^{(j)}_2,...,X^{(j)}_p\}$
\EndFor
\end{algorithmic}
\end{algorithm}

\textit{Lines 2-4}: partitions the input data. $\Theta^T$ is evenly split by columns into \textit{p} partitions, $X$ is evenly split by rows into \textit{q} partitions, and $R$ is split by rows and columns following the partition schemes of $X$ and $\Theta^T$.

\textit{Lines 5-7}: copies $\Theta^{T(i)}$ to GPU$_i$ ($1\leq i\leq p$), in parallel.

\textit{Lines 8-20}: loop over $\{X^{(1)}, X^{(2)},...,X^{(q)}\}$ and solve each $X^{(j)}$ partition in sequence ($1\leq j\leq q$). Given more GPUs, this sequential loop can further be parallelized.

\textit{Line 9-18}: parallel loop over $\{\Theta^{T(i)}\}$ ($1\leq i\leq p$) to solve $X^{(j)}$. Without sufficient number of GPUs, this \textbf{parallel for} loop can degrade to a \textbf{sequential} one.

\textit{Lines 11}: on GPU$_i$ ($1\leq i\leq p$), for each row $\textbf{x}_u$ in $X^{(j)}$, calculate the $A_u$ local to GPU$_i$ by only observing $\Theta^{T(i)}$ and $R^{(ij)}$:
\begin{align}
A_u^{i}=\sum\limits_{r_{uv}\neq0}^{GPU_i}(\theta_v \theta_v^T+\lambda I)
\end{align}

Similarly, we calculate the local $B_u$ matrix:
\begin{align}
B_u^{i}=\Theta^{T(i)}\cdot (R^{(ij)}_{u*})^T
\end{align}

The collection of all $A_u^{i}$s and $B_u^{i}$s on GPU$_i$ are denoted as $(A^{(ij)}, B^{(ij)})$.

\textit{Line 12}: a synchronization barrier to wait for all parfor threads to reach this step.

\textit{Lines 13-14}: evenly partition $A^{(ij)}$ and $B^{(ij)}$ by rows of $X^{(j)}$. That is, $A^{(ij)}$ on GPU$_i$ is evenly divided into $p$ portions:

\begin{center}
$A^{(ij)}_1$, $A^{(ij)}_2$, ..., $A^{(ij)}_p$
\end{center}

$B^{(ij)}$ is partitioned in the same manner into:
\begin{center}
$B^{(ij)}_1$, $B^{(ij)}_2$, ..., $B^{(ij)}_p$
\end{center}

\textit{Lines 15-16}: \textbf{parallel reduce} $p$ $A^{(ij)}$s and $B^{(ij)}$s into the global $A^{(j)}$ and $B^{(j)}$, on $p$ GPUs. GPU$_i$ takes care of the reduction of partition $i$ of all $A^{(kj)}$s ($1\leq k\leq p$). See Figure~\ref{fig:par-reduce} (a) for an example where $j=1$ and $p=4$: GPU$_1$ reduces $\{A^{(11)}_1, A^{(21)}_1, A^{(31)}_1, A^{(41)}_1\}$,  GPU$_2$ reduces $\{A^{(11)}_2, A^{(21)}_2, A^{(31)}_2, A^{(41)}_2\}$, and so on. $B^{(ij)}$s are reduced in the same manner.

\textit{Line 17}: solves the $p$ partitions concurrently on $p$ GPUs. GPU${_i}$ solves the local partition ($A^{(j)}_i,B^{(j)}_i$) it reduces in \textit{Lines 15-16}. 

\textit{Line 19}: obtain $X^{(j)}$ by collecting $p$ partitions $\{X^{(j)}_1,X^{(j)}_2,...,X^{(j)}_p\}$ on $p$ GPUs.

\subsection{Topology-aware parallel reduction to speed up SU-ALS}
\begin{figure}[tb]
\subfloat[One-phase parallel reduction.]{%
  \includegraphics[clip,width=0.9\columnwidth]{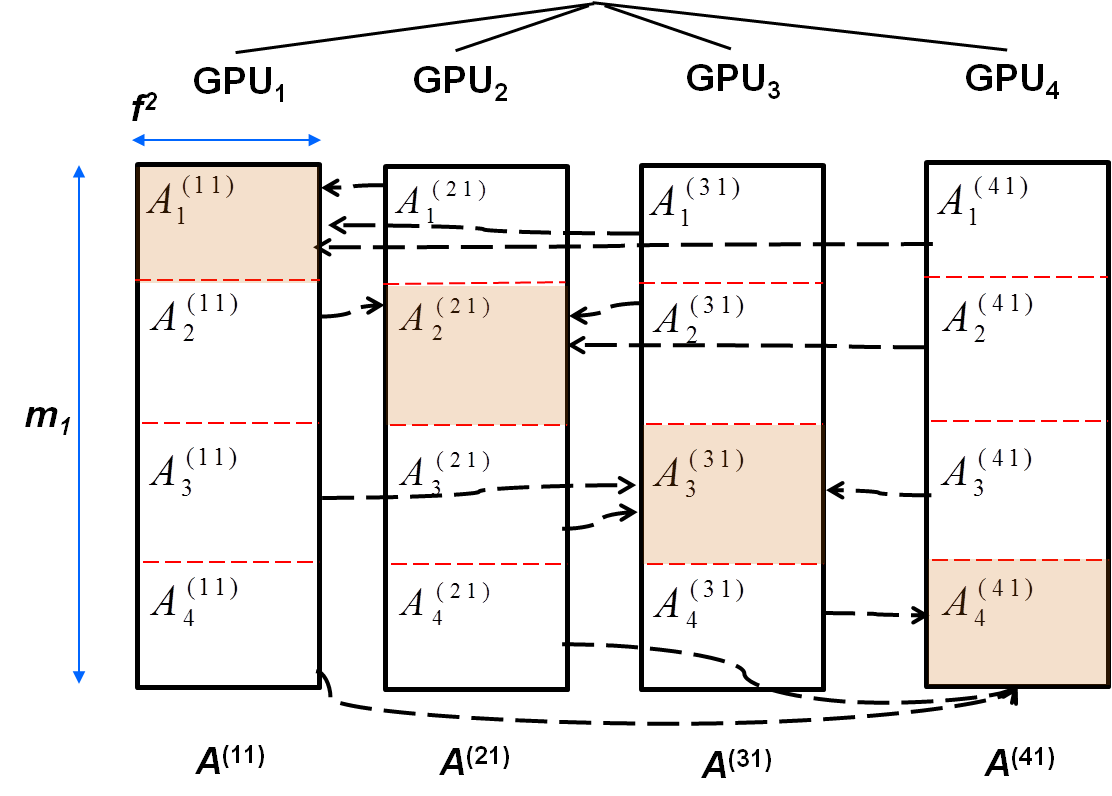}
}

\subfloat[Two-phase parallel reduction considering PCIe hierarchy: phase-1 (intra-socket) in dash lines; phase-2 (inter-socket) in solid lines.]{
  \includegraphics[clip,width=0.87\columnwidth]{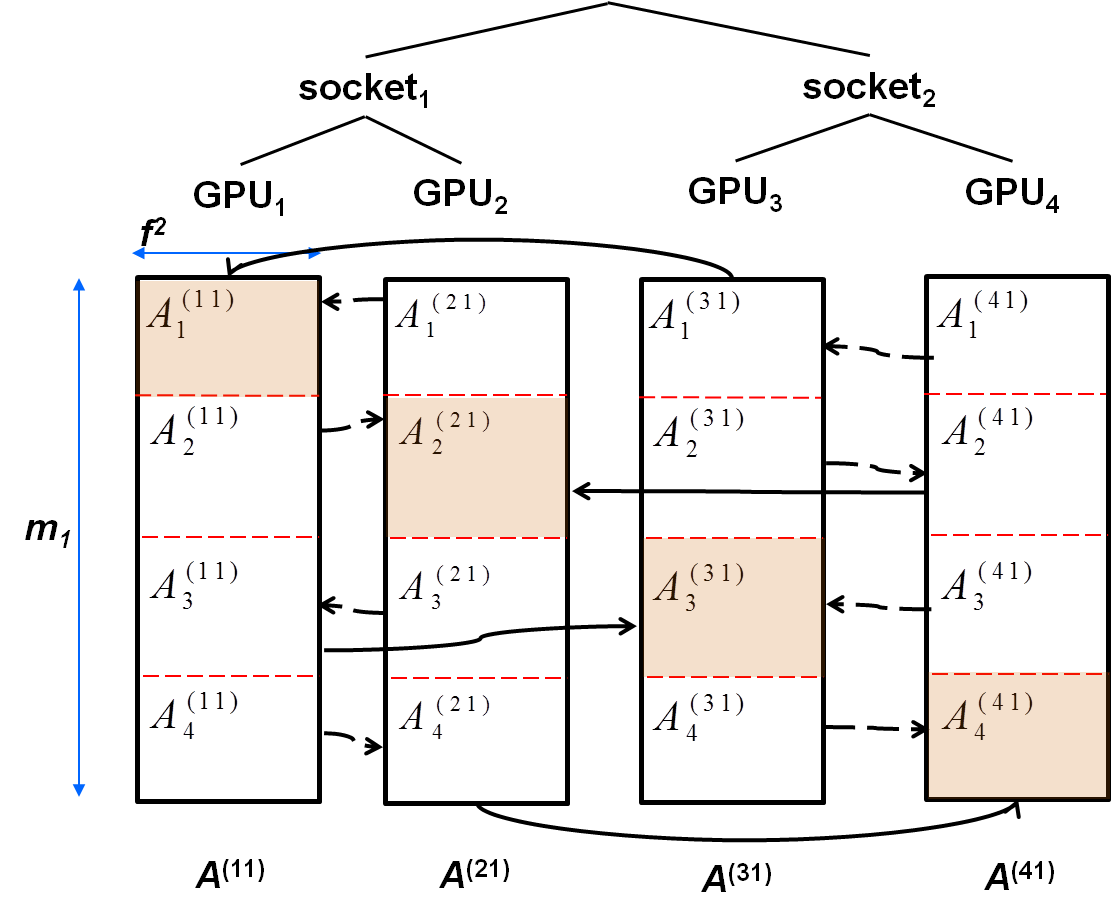}
}
\caption{Parallel reduce $A^{(ij)}$ in SU-ALS when $j=1$ and $p=4$. For $1\leq i\leq p$, on GPU$_i$, $A^{(ij)}$ is evenly partitioned into $p$ pieces: $A^{(ij)}_1, A^{(ij)}_2, ..., A^{(ij)}_p$. Afterward GPU$_i$ reduces all $A^{(kj)}_i$ across $p$ GPUs ($1\leq k\leq p$). This not only achieves parallel get\_hermitian\_x and batch\_solve, but also leverages cross-GPU bandwidth efficiently.}
\label{fig:par-reduce}
\end{figure}
\noindent\textbf{Parallel reduction.}
In Lines 13-17 of Algorithm~\ref{alg:su-als}, ($A^{(j)}, B^{(j)}$) could have been reduced in one GPU (say, GPU$_1$) and $X^{(j)}$ solved there. However, this simple approach fails to parallelize either data transfer or computation. Moreover, multiple GPUs on a machine are usually connected through a PCIe bus. PCIe channels are full-duplex, meaning that data transfer in both directions can happen simultaneously without affecting each other. To leverage the bandwidth in both directions, we develop a parallel reduction scheme that evenly utilizes both incoming and outgoing channels of all GPUs, as shown in Figure~\ref{fig:par-reduce} (a). Experiment on Hugewiki data set shows that this optimization is 1.7x as fast compared with the reducing-by-one-GPU approach. After this parallel reduction, \textsl{batch\_solve} begins on $p$ GPUs in parallel.

\noindent\textbf{Topology-aware parallel reduction.}
Figure~\ref{fig:par-reduce} (a) assumes a flat interconnection where all GPUs directly connect to a PCIe root. This assumption may not always hold. For example, in a two-socket machine with four GPUs, a typical configuration is that every two GPUs connect to one socket. Communications between the two GPUs in the same socket still go though the local PCIe bus, while communications between GPUs in different sockets go through the inter-socket connection. In this case, intra-socket transfers enjoy zero-copy and faster duplex PCIe channel, compared with inter-socket transfers. In such a topology, the scheme shown in Figure~\ref{fig:par-reduce} (a) is not optimal.

Based on the GPU connection topology, we design a two-phase parallel reduction scheme shown in Figure~\ref{fig:par-reduce} (b). In this scheme, each partition is first reduced intra socket (see the dash line). Afterward, the partial, intra-socket reduction results are moved across socket and generate the final reduction result (the solid line). Experiments show that this two-phase scheme enjoys an additional 1.5x speedup compared with the one-phase scheme shown in Figure~\ref{fig:par-reduce} (a).

\subsection{How to partition?}
%
%

Assume a single GPU's memory capacity is $C$. According to Algorithm~\ref{alg:su-als}, one GPU needs to hold $X^{(j)}$, $\Theta^{(i)}$, $R^{(ij)}$, $A^{(j)}$, and $B^{(j)}$. Therefore the choices of $p$ and $q$ are subject to \eqref{eq:partition}.

\begin{align}
\label{eq:partition}
 \frac{m\times f}{q} + \frac{n\times f}{p} + |R^{(ij)}| + \frac{m}{q}\times f^2 + \frac{m}{q}\times f + \epsilon < C
\end{align}

$\epsilon$ is a headroom space for miscellaneous small variables. In practice, when $C=12$ GB we choose $\epsilon=500$ MB.
 
Here are some best practices in choosing $p$ and $q$:
\begin{enumerate}
	\item If $p=1$ can satisfy \eqref{eq:partition}, you can solve $X$ in a single GPU in sequential batches. In this case SU-ALS is equivalent to MO-ALS.
	\item When $q$ increases and $p=1$ satisfies \eqref{eq:partition}, $q$ should not increase any more. At this time there is already no need to further partition $X$. 
	\item We usually start from $p$ such that $\displaystyle{\frac{n\times f}{p}} \approx \displaystyle{\frac{C}{2}} $, and then choose the smallest $q$ that satisfies \eqref{eq:partition}. 
\end{enumerate}
\subsection{Implementation of cuMF}
This section describes selected details of cuMF. CuMF is implemented in C, using CUDA 7.0 and GCC OpenMP v3.0. It has circa 6,000 lines of code. 

\textbf{Out-of-core computation.} As seen in Figure~\ref{fig:scale} and Table~\ref{tbl:dataset}, rating and feature matrices can both have 100 billion entries. This goes far beyond the host and device memory limit. For such out-of-core problems, cuMF first generate a partition scheme, planning which partition to send to which GPU in what order. With this knowledge in advance, cuMF uses separate CPU threads to preload data from disk to host memory, and separate CUDA streams to preload from host memory to GPU memory. By this proactive and asynchronous data loading, we manage to handle out-of-core problems with close-to-zero data loading time except for the first load.  

\textbf{Elasticity to resources.}
Algorithm~\ref{alg:su-als} is generic enough to cover many deployment scenarios where the number of GPUs are fewer or more than $p$ or $q$. With more GPUs, the sequential \textbf{for} at \textit{Line 8} can be parallelized; with fewer GPUs, the \textbf{parfor} at \textit{Line 9} can be turned into a sequential for. This is similar to how MapReduce deals with resource elasticity: when there are fewer/more parallel tasks compared with task slots, tasks will be executed in fewer/more waves. By this design cuMF is able to solve ALS of any size.

\textbf{Fault tolerance.} Handling machine failure is straightforward in cuMF which uses a single machine. During ALS execution we asynchronously checkpoint $X$ and $\Theta$ generated from the latest iteration, into a connected parallel file system. When the machine fails, the latest $X$ or $\Theta$ (whichever is more recent) is used to restart ALS.

%% file: exp.tex
\section{Experiments}
\label{sec:exp}
This section reports the performance evaluations on cuMF. We compare cuMF with multi-core solutions libMF \cite{libmf-13} and NOMAD \cite{nomad14}. We also compare with distributed solutions including NOMAD (on multi-nodes), Factorbird \cite{factorbird14}, Spark ALS \cite{sparkals14}, and a Giraph based solution from Facebook \cite{facebook15}. We select these solutions because they either perform better than earlier studies ~\cite{DSGD-kdd11, DBLP:conf/icdm/TeflioudiMG12, sparkler13, fastccd, ccd++-icdm12}, or are able to handle large data sets. Because none of existing GPU-based solutions \cite{gpu-rbm, Zastrau:2012:SGD} can tackle big data sets, we do not compare with their results.

The goals of our experiments are to provide key insights on the following questions:
\begin{enumerate}
	\item how would cuMF on a single GPU compare with highly optimized multi-core methods, such as libMF and NOMAD, on medium-size problems? (Section \ref{sec:exp-one-gpu})
	\item are the memory optimization done by MO-ALS effective? (Section \ref{sec:exp-mo})
    \item is SU-ALS scalable with multiple GPUs? (Section \ref{sec:exp-scalability})
	\item with four GPUs on one machine, how would cuMF compare with multi-node methods on large-size problems? (Section \ref{sec:exp-xlarge})
\end{enumerate}
\subsection{Experiment setting}
\textbf{Data Sets}. We use three public data sets, i.e., Netflix \cite{netflix08}, YahooMusic \cite{kdd-cup-yahoomusic-11} and Hugewiki~\cite{libmf-13} to measure the convergence speed. For large-size problems, we synthesize the data sets used by SparkALS~\cite{sparkals14}, Factorbird~\cite{factorbird14} and Facebook~\cite{facebook15}. For these three systems, we compare the per iteration latency because their convergence speed are not reported. We also synthesize a data set to the size that is beyond any previous attempts. That is, we use the rating matrix of the Facebook data set, with an enlarged $f$ of 100 from the original 16. Characteristics of these data sets are shown in Table \ref{tbl:dataset}.

\textbf{Hardware}. Unless otherwise mentioned, we use one to four Nvidia Titan X GPUs, each with 3072 CUDA cores and 12 GB memory, on one machine. The machine is with two Intel Xeon E5 CPUs, 256 GB RAM, and the GPFS \cite{gpfs} as the file system.  

\textbf{Parameters}. The $f$ and $\lambda$ values for each data set are given in Table \ref{tbl:dataset}. Feature matrices are initiated with random numbers in $[0,1]$. We focus on the speed and scalability of cuMF, and therefore did not spend much effort in hyper-parameter tuning to achieve the best accuracy.

\textbf{Evaluation}. For Netflix, YahooMusic and Hugewiki, we evaluate the root-mean-square-error (RMSE) on test set. Performance of libMF and NOMAD is obtained from \cite{libmf-13, nomad14}. For SparkALS, Factorbird and Facebook, since the data is synthetic and no test RMSE is reported, we compare the per iteration run time.
\begin{table}
\centering
\caption{Data sets}
\label{tbl:dataset}
\begin{tabular}{|c|c|c|c|c|c|} \hline
\textbf{Data Set}	& \textbf{$m$} & \textbf{$n$}	& \textbf{$N_z$} & \textbf{$f$} &\textbf{$\lambda$} \\\hline
Netflix &480,189&17,770&99M&100&0.05\\ \hline
YahooMusic&1,000,990&624,961&252.8M&100&1.4\\ \hline
Hugewiki&50,082,603&39,780&3.1B&100&0.05 \\ \hline
\hline
SparkALS&660M&2.4M&3.5B&10&0.05\\ \hline
Factorbird&229M&195M&38.5B&5&0.05\\ \hline
Facebook&1B&48M&112B&16&0.05\\ \hline
\textbf{cuMF}&1B&48M&112B&100&0.05\\ \hline
\end{tabular}
\end{table}
\subsection{MO-ALS on a single GPU}
\label{sec:exp-one-gpu}
\begin{figure} %
    \centering
    \subfloat[Netflix]{%
    \includegraphics[width=0.25\textwidth]{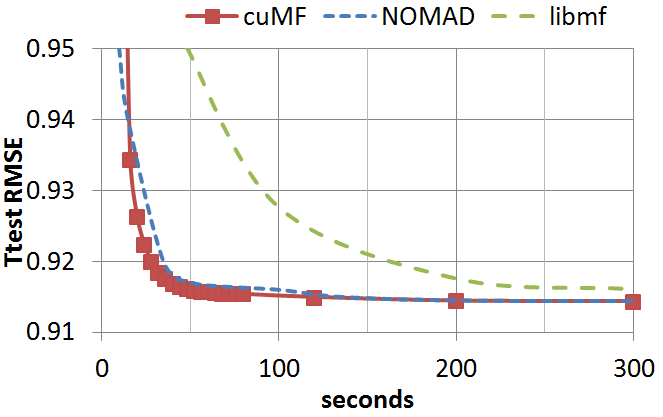}} %
    \subfloat[YahooMusic]{%
    \includegraphics[width=0.25\textwidth]{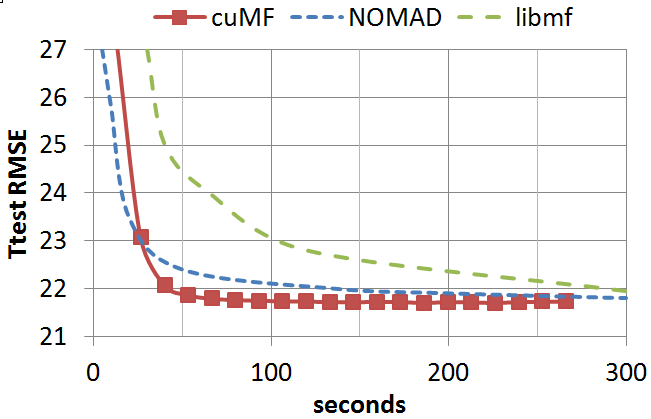}} %
    \caption{%
    \label{fig:netflix-yahoo-perf} %
    Test RMSE convergence speed in terms of number of iterations: cuMF (with one GPU), NOMAD and libMF (both with 30 CPU cores).}
\end{figure}

We run cuMF on one GPU, measure the test RMSE w.r.t. training time, and compare with NOMAD and libMF on one machine with 30 cores \cite{nomad14}. We choose these two for comparison because they are among the fastest multi-core solutions. In Figure~\ref{fig:netflix-yahoo-perf}, on both Netflix and YahooMusic, cuMF performs slightly worse than NOMAD at the beginning but slightly better later, and constantly faster than libMF. CuMF use ALS where each iteration takes much longer than SGD based methods. This makes it slower at the beginning. Nevertheless cuMF catches up quickly and outperforms soon afterward. 

\subsection{Benefit of using register and texture memory in MO-ALS}
\begin{figure} %
    \centering
    \subfloat[Netflix]{%
    \includegraphics[width=0.25\textwidth]{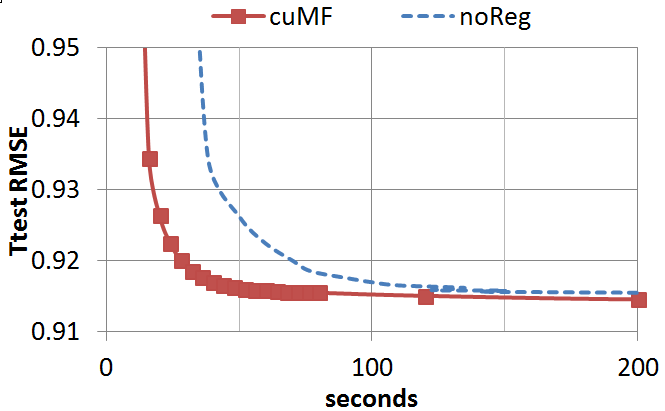}} %
    \subfloat[YahooMusic]{%
    \includegraphics[width=0.25\textwidth]{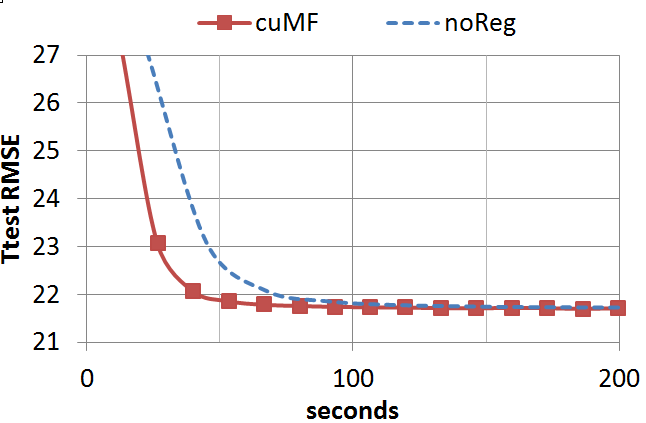}} %
    \caption{%
    \label{fig:netflix-yahoo-noReg} %
    The convergence speed of cuMF, with or without aggressively using registers on one GPU.}
\end{figure}

We first measure the benefit of aggressively using registers in MO-ALS. Figure \ref{fig:netflix-yahoo-noReg} compares cuMF's performance, with or without using register memory to aggregate $A_u$, on one GPU. On Netflix data, cuMF converges 2.5 times as slow (75 seconds vs. 30 seconds when RMSE reaches 0.92) without using registers. The result strongly supports the idea of aggressively using registers. Among all optimizations done in MO-ALS, using registers for $A_u$ brings the greatest performance gain. Without using the registers, cuMF converges 1.7 times as slow on YahooMusic. YahooMusic has a smaller performance degradation without using registers than Netflix. This is because its rating matrix is more sparse. As a result, its \textsc{Get\_Hermitian\_X()} is less heavy-duty and occupy a smaller percentage of the overall run time.

\begin{figure} %
    \centering
    \subfloat[Netflix]{%
    \includegraphics[width=0.25\textwidth]{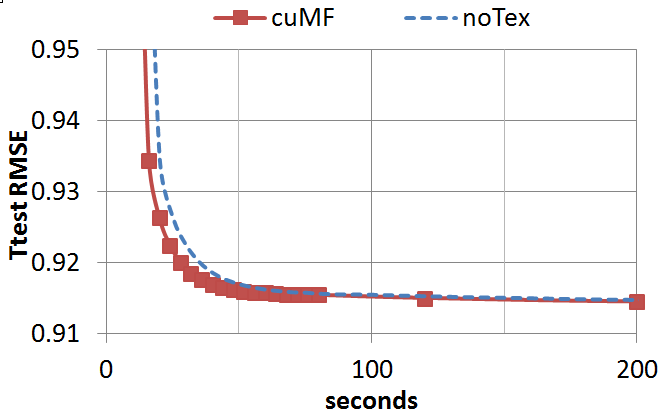}} %
    \subfloat[YahooMusic]{%
    \includegraphics[width=0.25\textwidth]{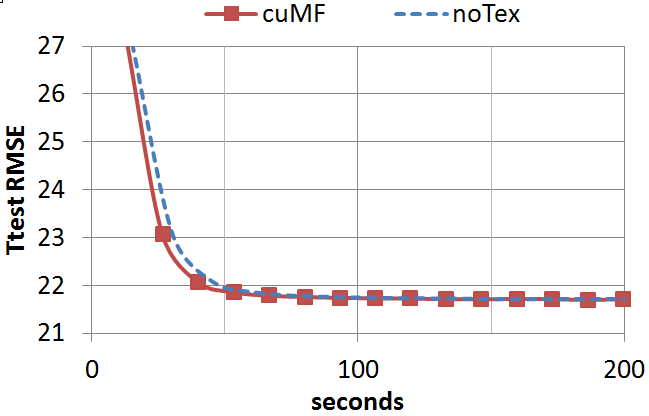}} %
    \caption{%
    \label{fig:netflix-yahoo-noTex} %
    The convergence speed of cuMF, with or without texture memory on one GPU.}
\end{figure}

Figure \ref{fig:netflix-yahoo-noTex} compares cuMF's performance with or without using texture memory. Using texture memory, the convergence speed is 25\% to 35\% faster. The reason for the gain is due to the fact that Algorithm \ref{alg:mo-als} updates $\Theta$ and $X$ in an alternating manner, i.e., $\Theta$ is read-only when updating $X$, and $X$ is read-only when updating $\Theta$. This feature enables us to leverage the read-only texture memory in GPU to speed up memory access. Since YahooMusic data is more sparse, the penalty of not using texture memory is also smaller. 
\label{sec:exp-mo}

\subsection{Scalability of SU-ALS on multiple GPUs}
\label{sec:exp-scalability}
\begin{figure} %
    \centering
    \subfloat[Netflix]{%
    \includegraphics[width=0.25\textwidth]{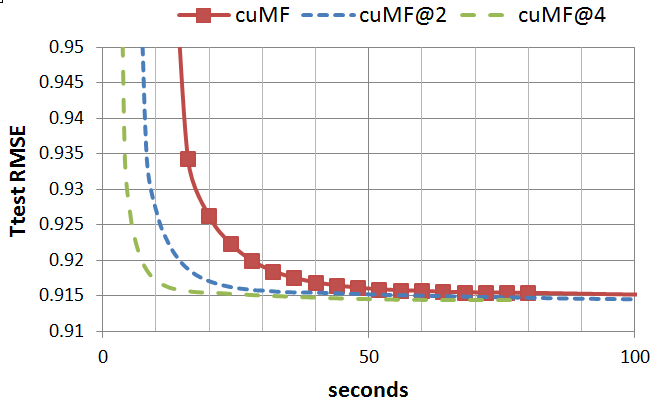}} %
    \subfloat[YahooMusic]{%
    \includegraphics[width=0.25\textwidth]{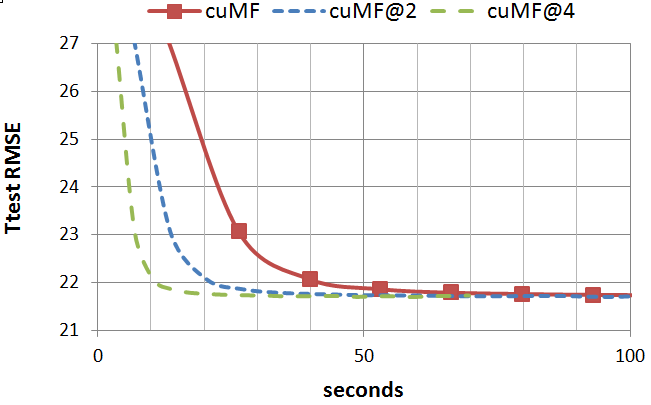}} %
    \caption{%
    \label{fig:netflix-yahoo-124} %
    The convergence speed of cuMF on one, two, and four GPUs.}
\end{figure}
This section first studies how a problem with the fixed size data set can be accelerated with multiple GPUs. In both Netflix and YahooMusic, $X$ and $\Theta$ can both fit on one GPU. As a result only model parallelism is needed. We run Netflix and YahooMusic data on one, two and four GPUs, respectively, on one machine. As seen from Figure~\ref{fig:netflix-yahoo-124}, close-to-linear speedup is achieved. For example, the speedup is 3.8x when using four GPUs, measured at RMSE 0.92. Detailed profiling shows that, the very small overhead mainly comes from PCIe IO contention when multiple GPUs read from host memory simultaneously. 

In contrast, NOMAD observed a sub-linear speedup on certain data sets, due to cache locality effects and communication overhead \cite{nomad14}. CuMF achieves better scalability due to the optimized memory access and inter-GPU communication. An advantage of cuMF is that, it consolidates massive computation on a single machine, so that it only uses PCIe connections which are faster than any existing network. 

\begin{figure}
\center{\includegraphics[width=0.35\textwidth]
        {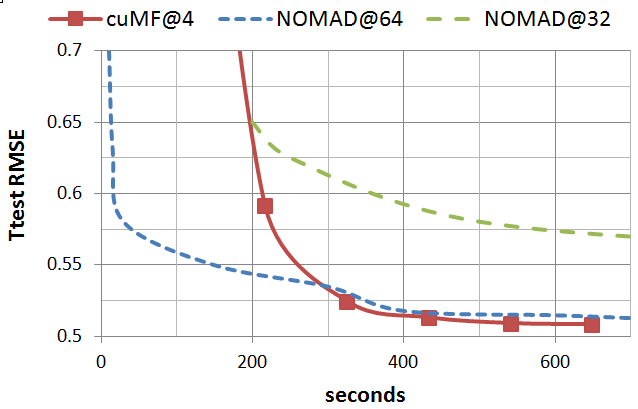}}
 \caption{CuMF@4GPU, vs. NOMAD on a 64-node HPC cluster and a 32-node AWS cluster, with Hugewiki data. CuMF converges similar to NOMAD with 64 nodes, and 10x as fast as NOMAD with 32 nodes.}
 \label{fig:hugewiki}
\end{figure}

We also tested Hugewiki data on four GPUs. We compare with multi-node NOMAD (on 64-node HPC cluster and 32-node AWS cluster) because it outperforms DSGD~\cite{DSGD-kdd11} and DSGD++~\cite{DBLP:conf/icdm/TeflioudiMG12}.
Hugewiki is a relatively large data set where $m\approx50$M, $n\approx40$K, and $N_z\approx 3$B. When using $X$ to solve $\Theta$, $X$ is too big to fit on one GPU. According to Algorithm~\ref{alg:su-als} we partition $X$ evenly into four GPUs and apply data parallelism. We use the two-phase parallel reduction scheme shown in Figure \ref{fig:par-reduce} (b), because our machine has two sockets each connecting to two GPUs. With all the intra- and inter-GPU optimizations, cuMF performs slightly better than NOMAD on a 64-node HPC cluster (again, with a slower start), and much better than NOMAD on a 32-node AWS cluster, as shown in Figure ~\ref{fig:hugewiki}. This result is very impressive, as a 64-node HPC cluster is outperformed by only one node plus four GPUs. This indicates that cuMF brings a big saving in infrastructure and management cost. 

\subsection{Solve extremely large-scale problems}
\label{sec:exp-xlarge}
\begin{figure}
\center{\includegraphics[width=0.35\textwidth]
        {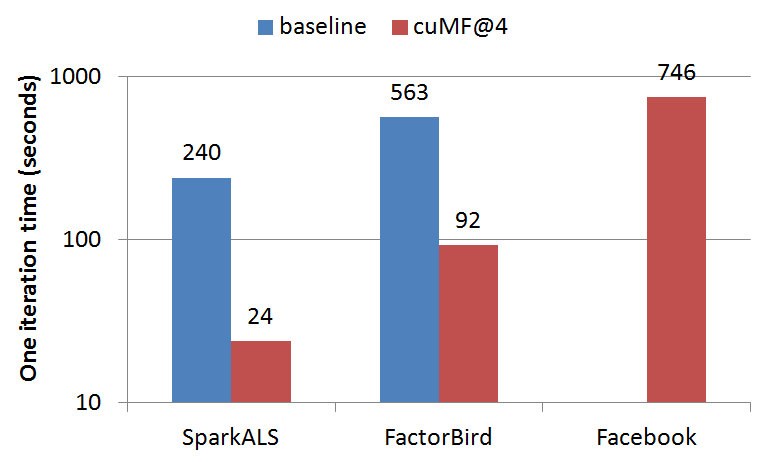}}
 \caption{CuMF@4GPU on three very large data sets, compared with the their original implementations as baselines.}
\label{fig:verylarge}
\end{figure}
We conduct experiments on three extremely large problems. In this experiment we use four Nvidia GK210 cards on one machine. Each card is with 2496 CUDA cores (slightly fewer than Titan X) and 12 GB memory, and every two cards are encapsulated as one K80 GPU. 

The results for the following experiments are shown in Figure~\ref{fig:verylarge}.
SparkALS~\cite{sparkals14} is a benchmark of Spark MLlib ALS.
Its rating matrix is from the 100-by-1 duplication of the \textit{Amazon Reviews}~\cite{amazonreviews} data. It uses 50$\times$m3.2xlarge AWS nodes with Spark MLlib 1.1, and takes 240 seconds per ALS iteration. We synthesize the data in the same way as~\cite{sparkals14}, apply model parallelism solving $X$, and apply data parallelism solving $\Theta$. CuMF with four GPUs completes one iteration in 24 seconds, which is \textbf{ten times as fast} as SparkALS. 

Factorbird~\cite{factorbird14} is a parameter server system for MF. It trains a data set ($m=229$M, $n=195$M, $f=5$, and $N_z=38.5$B) on a cluster of 50 nodes. We synthesize the data using the method described in~\cite{DBLP:conf/icdm/TeflioudiMG12}. We use only model parallelism in solving $X$ and $\Theta$ because they both fit into one GPU. CuMF with four GPUs completes one iteration in 92 seconds. Factorbird needs 563 seconds per iteration, and with SGD it may need more iterations than ALS.

Facebook~\cite{facebook15} recently revealed that its MF system deals with 1 billion users, millions of items and over 100 billion ratings. Given this hint we did a 160-by-20 duplication of the Amazon Review data, yielding a data set with $m=1056$M, $n=48$M, $f=16$, and $N_z=112$B. We use data parallelism to solve both $X$ and $\Theta$. Especially, when solving $\Theta$, because $X$ is huge ($1056$M$\times16$ floats) and cannot fit on 4 GPUs, we change the \textbf{parfor} in Line 9-18 of Algorithm~\ref{alg:su-als} into a \textbf{sequential for} with many batches. By doing this, cuMF completes one ALS iteration in 746 seconds. \cite{facebook15} does not report its speed on 50 Giraph workers, but we believe cuMF is competitive given the size of the problem and the low cost of one machine with GPUs. We further try a larger $f=100$, and cuMF completes one iteration in 3.8 hours. To the best of our knowledge, this is by far the largest matrix factorization problem ever reported in literature.

As a summary, on two extremely large data sets, CuMF with four GPUs significantly outperforms the original distributed implementations. CuMF is also able to factorize the largest collaborative filtering matrix ever reported.

%% file: relatedwork.tex
\section{Related Work}
SGD, Coordinate Gradient Descent (CGD) and ALS are the three main algorithms for MF. This section firstly reviews the three algorithms and then the methods to parallelize them. Subsequently, we review GPU-based MF solutions.
\subsection{MF algorithms}
SGD based algorithms \cite{mf-computer09} have been often applied to matrix factorization. 
SGD handles large scale problems by splitting the rating matrix into blocks along with sophisticated conflict-avoiding updates. CGD based algorithms update along one coordinate direction in each iteration. \cite{fastccd} improved the default cyclic CGD scheme by prioritizing the more important coordinates. ALS algorithms \cite{netflix08, als-10} have advantages in easy to parallelize, converging in fewer iterations, and dealing with non-sparse rating matrices \cite{mf-computer09}. CuMF is based on ALS.
\subsection{Parallel computing paradigms}
\textbf{Parallel SGD.} SGD has been parallelized in environments including multi-core \cite{libmf-13}, multi-node MPI \cite{DBLP:conf/icdm/TeflioudiMG12, nomad14}, MapReduce \cite{DSGD-kdd11, sparkler13} and parameter-server \cite{factorbird14, Xing2014-PS}. These studies are inspired by HOGWILD! \cite{hogwild-nips11}, which shows how to avoid expensive memory locking in memory sharing systems for some optimization problems with sparse updates. These methods partition the rating matrix into blocks with no overlapping rows or columns, and work on these blocks in parallel. They also use asynchronous communication, overlapping of communication and computation, and shared memory to achieve further speedup.

LibMF \cite{libmf-13} is a very efficient SGD based library for matrix factorization on multi-cores. It has out performed nearly all other approaches on a 12-core machine. However, our experimental results show that libMF stops scaling beyond 16 cores, similar to the observation of \cite{dcMF15}. Moreover, libMF is a single-machine implementation, which limits its capability to solve large-scale problems. NOMAD \cite{nomad14} extends the idea of block partitioning, adding the capability to release a portion of a block to another thread before its full completion. It performs similar to libMF on a single machine, and can scale out to a 64-node HPC cluster.

\textbf{Parameter Server with SGD.} More recently, the idea of ``parameter server" \cite{Smola2014-PS, Xing2014-PS} emerges for extremely large-scale machine learning problems. In this paradigm, the \textit{server nodes} store parameters, while the \textit{worker nodes} store training data and compute on them. The parameter-server framework manages asynchronous communication between nodes, flexible consistency models, elastic scalability, and fault tolerance. Following this idea, Petuum \cite{Xing2014-PS} runs Netflix data 
on a 512 cores cluster using SGD. 
Factorbird \cite{factorbird14} is a parameter server specifically implemented for matrix factorization, also based on SGD.  

\textbf{Parallel CGD.} CCD++ \cite{ccd++-icdm12} performs sequential updates on one row of the decomposed matrix while fixing other variables. CCD++ has lower time complexity but makes less progress per iteration, compared with ALS. In practice, CCD++ behaves well in the early stage of optimization, but then becomes slower than libMF.

\textbf{Parallel ALS.} As discussed in Section~\ref{sec:als-challenges}, PALS \cite{netflix08} and SparkALS \cite{sparkmllib15} parallelize ALS by feature matrix replication and partial replication, respectively. These approaches does not work when feature matrices get extremely large. Facebook \cite{facebook15} tackles this issue by feeding a feature matrix in parts to a node. For example, when solving $X$, $X$ is partitioned disjointedly across nodes; $\Theta$ is also partitioned and rotated across the same set of nodes. 
When a $\Theta$ partition $\Theta^{(j)}$ meets $X$ partition $X^{(i)}$, $X^{(i)}$ is updated by observing $\Theta^{(j)}$; $X^{(i)}$ completes an iteration of update after it meets all $\Theta^{(j)}$s. 
This is somewhat similar to SU-ALS but SU-ALS does not use rotation, 
as GPUs do not have sufficient memory to do rotation.

GraphLab \cite{graphlab12} implements ALS in such a way that when $\Theta$ is big, it is distributed among multiple machines. When updating a $\textbf{x}_u$ in a node, all needed $\theta_v$s are fetched on-the-fly from all nodes. This involves a lot of cross-node traffic and puts a high requirement on network bandwidth.

\subsection{GPU approaches}
\cite{gpu-rbm} employs GPU-based restricted Boltzmann machine for collaborative filtering, which gives relative performance compared with a CPU implementation on Netflix data. \cite{Zastrau:2012:SGD} implements both SGD and ALS on GPU to solve MF. It uses a mini-batch-based and sequential version of SGD, and a variant of ALS that adjusts (rather than re-calculates) the inverse of the Hermitian matrices in each iteration. They neither optimize the memory access to fully utilize GPU's compute power, nor scale to multiple GPUs to handle large-scale problems. 

Compared with CPU-based approaches, cuMF has better performance with a fraction of hardware resources. Compared with GPU-based approaches, our optimization in memory access and parallelism yields higher performance and scalability.


%% file: conclusion.tex
\section{Conclusion}


Advances in GPU computing opens new possibilities to accelerate high performance parallel and large scale distributed applications. GPUs enable us to consolidate huge compute power and memory bandwidth on one or few machines, which may reduce the demand for big distributed clusters. This scale-up approach provides an alternative to the scale-out systems in distributed applications. Evidently, cuMF using a single machine with GPUs is faster and cheaper to solve matrix factorization, compared with distributed CPU systems. CuMF achieves this by optimizing memory access, combining data and model parallelism, and applying topology-aware parallel reduction.

In future work we plan to extend cuMF to deal with other sparse problems such as graph algorithms \cite{hpdc2014gpugraph}, and use it to accelerate Hadoop/Spark framework~\cite{hpdc2015gpu}.


%% file: cuMF.bbl
\begin{thebibliography}{10}

\bibitem{gpu-rbm}
X.~Cai, Z.~Xu, G.~Lai, C.~Wu, and X.~Lin.
\newblock {GPU}-accelerated restricted boltzmann machine for collaborative
  filtering.
\newblock In {\em ICA3PP}, pages 303--316, 2012.

\bibitem{DBLP:conf/emnlp/CannyHK13}
J.~Canny, D.~L.~W. Hall, and D.~Klein.
\newblock A multi-teraflop constituency parser using {GPUs}.
\newblock In {\em EMNLP}, pages 1898--1907, 2013.

\bibitem{costshpc2013}
A.~Coates, B.~Huval, T.~Wang, D.~Wu, B.~Catanzaro, and N.~Andrew.
\newblock {Deep learning with COTS HPC systems}.
\newblock In {\em ICML}, pages 1337--1345, 2013.

\bibitem{Xing2014-PS}
H.~Cui, J.~Cipar, Q.~Ho, J.~K. Kim, S.~Lee, A.~Kumar, J.~Wei, W.~Dai, G.~R.
  Ganger, P.~B. Gibbons, G.~A. Gibson, and E.~P. Xing.
\newblock Exploiting bounded staleness to speed up big data analytics.
\newblock In {\em USENIX ATC}, pages 37--48, 2014.

\bibitem{jeffdean-dnn-nips12}
J.~Dean, G.~S. Corrado, R.~Monga, K.~Chen, M.~Devin, Q.~V. Le, M.~Z. Mao,
  M.~Ranzato, A.~Senior, P.~Tucker, K.~Yang, and A.~Y. Ng.
\newblock Large scale distributed deep networks.
\newblock In {\em NIPS}, pages 1223--1231, 2012.

\bibitem{kdd-cup-yahoomusic-11}
G.~Dror, N.~Koenigstein, Y.~Koren, and M.~Weimer.
\newblock {The Yahoo! Music Dataset and KDD-Cup '11}.
\newblock In {\em {KDD} Cup 2011 competition}, 2012.

\bibitem{DSGD-kdd11}
R.~Gemulla, E.~Nijkamp, P.~J. Haas, and Y.~Sismanis.
\newblock Large-scale matrix factorization with distributed stochastic gradient
  descent.
\newblock In {\em KDD}, pages 69--77, 2011.

\bibitem{hennessy2011computer}
J.~L. Hennessy and D.~A. Patterson.
\newblock {\em Computer architecture: a quantitative approach}.
\newblock Elsevier, 2011.

\bibitem{fastccd}
C.-J. Hsieh and I.~S. Dhillon.
\newblock Fast coordinate descent methods with variable selection for
  non-negative matrix factorization.
\newblock In {\em KDD}, pages 1064--1072, 2011.

\bibitem{gpfs}
{IBM}.
\newblock {General Parallel Filesystem}.
\newblock
  \url{http://www-01.ibm.com/support/knowledgecenter/?lang=en#!/SSFKCN_4.1.0.4/gpfs.v4r104_welcome.html},
  2014.

\bibitem{facebook15}
M.~Kabiljo and A.~Ilic.
\newblock {Recommending items to more than a billion people}.
\newblock \url{https://code.facebook.com/posts/861999383875667}, 2015.
\newblock [Online; accessed 17-Aug-2015].

\bibitem{hpdc2014gpugraph}
F.~Khorasani, K.~Vora, R.~Gupta, and L.~N. Bhuyan.
\newblock Cusha: Vertex-centric graph processing on gpus.
\newblock In {\em HPDC}, pages 239--252, 2014.

\bibitem{mf-computer09}
Y.~Koren, R.~M. Bell, and C.~Volinsky.
\newblock Matrix factorization techniques for recommender systems.
\newblock {\em Computer}, 42(8):30--37, 2009.

\bibitem{GeMTC2014}
S.~J. Krieder, J.~M. Wozniak, T.~Armstrong, M.~Wilde, D.~S. Katz, B.~Grimmer,
  I.~T. Foster, and I.~Raicu.
\newblock Design and evaluation of the gemtc framework for gpu-enabled
  many-task computing.
\newblock In {\em HPDC}, pages 153--164, 2014.

\bibitem{sparkler13}
B.~Li, S.~Tata, and Y.~Sismanis.
\newblock Sparkler: Supporting large-scale matrix factorization.
\newblock In {\em EDBT}, pages 625--636, 2013.

\bibitem{Smola2014-PS}
M.~Li, D.~G. Andersen, J.~W. Park, A.~J. Smola, A.~Ahmed, V.~Josifovski,
  J.~Long, E.~J. Shekita, and B.-Y. Su.
\newblock Scaling distributed machine learning with the parameter server.
\newblock In {\em OSDI}, pages 583--598, 2014.

\bibitem{graphlab12}
Y.~Low, D.~Bickson, J.~Gonzalez, C.~Guestrin, A.~Kyrola, and J.~M. Hellerstein.
\newblock Distributed {GraphLab}: a framework for machine learning and data
  mining in the cloud.
\newblock In {\em VLDB}, pages 716--727, 2012.

\bibitem{sparkmllib15}
X.~Meng, J.~K. Bradley, B.~Yavuz, E.~R. Sparks, S.~Venkataraman, D.~Liu,
  J.~Freeman, D.~B. Tsai, M.~Amde, S.~Owen, D.~Xin, R.~Xin, M.~J. Franklin,
  R.~Zadeh, M.~Zaharia, and A.~Talwalkar.
\newblock {MLlib: Machine Learning in Apache Spark}.
\newblock {\em CoRR}, abs/1505.06807, 2015.

\bibitem{dcMF15}
Y.~Nishioka and K.~Taura.
\newblock Scalable task-parallel sgd on matrix factorization in multicore
  architectures.
\newblock In {\em ParLearning}, 2015.

\bibitem{hogwild-nips11}
F.~Niu, B.~Recht, C.~Re, and S.~J. Wright.
\newblock {HOGWILD!}: A lock-free approach to parallelizing stochastic gradient
  descent.
\newblock In {\em NIPS}, pages 693--701, 2011.

\bibitem{cublas}
{Nvidia}.
\newblock {cuBLAS}.
\newblock \url{http://docs.nvidia.com/cuda/cublas/}, 2015.
\newblock [Online; accessed 17-Aug-2015].

\bibitem{cusparse}
{Nvidia}.
\newblock {cuSPARSE}.
\newblock \url{http://docs.nvidia.com/cuda/cusparse/#cusparse-lt-t-gt-csrmm2},
  2015.
\newblock [Online; accessed 4-Aug-2015].

\bibitem{pennington2014glove}
J.~Pennington, R.~Socher, and C.~D. Manning.
\newblock Glove: Global vectors for word representation.
\newblock In {\em EMNLP}, pages 1532--1543, 2014.

\bibitem{als-10}
I.~Pillaszy, D.~Zibriczky, and D.~Tikk.
\newblock Fast {ALS}-based matrix factorization for explicit and implicit
  feedback datasets.
\newblock In {\em {RecSys}}, pages 71--78, 2010.

\bibitem{flink15}
T.~Rohrmann.
\newblock {How to factorize a 700 GB matrix with Apache Flink}.
\newblock
  \url{http://data-artisans.com/how-to-factorize-a-700-gb-matrix-with-apache-flink/},
  2015.
\newblock [Online; accessed 15-Aug-2015].

\bibitem{optimizecuda08}
S.~Ryoo, C.~I. Rodrigues, S.~S. Baghsorkhi, S.~S. Stone, D.~B. Kirk, and
  W.-m.~W. Hwu.
\newblock Optimization principles and application performance evaluation of a
  multithreaded {GPU Using CUDA}.
\newblock In {\em PPoPP}, pages 73--82, 2008.

\bibitem{hpdc2015gpu}
A.~Sabne, P.~Sakdhnagool, and R.~Eigenmann.
\newblock Heterodoop: A mapreduce programming system for accelerator clusters.
\newblock In {\em HPDC}, pages 235--246, 2015.

\bibitem{factorbird14}
S.~Schelter, V.~Satuluri, and R.~B. Zadeh.
\newblock Factorbird-a parameter server approach to distributed matrix
  factorization.
\newblock In {\em NIPS Workshop on Distributed Matrix Computations}, 2014.

\bibitem{amazonreviews}
{Stanford SNAP Lab}.
\newblock {Web data: Amazon reviews}.
\newblock \url{https://snap.stanford.edu/data/web-Amazon.html}, 2015.
\newblock [Online; accessed 18-Aug-2015].

\bibitem{DBLP:conf/icdm/TeflioudiMG12}
C.~Teflioudi, F.~Makari, and R.~Gemulla.
\newblock Distributed matrix completion.
\newblock In {\em ICDM}, pages 655--664, 2012.

\bibitem{sparkals14}
B.~Yavuz, X.~Meng, and R.~Xin.
\newblock {Scalable Collaborative Filtering with Spark MLlib}.
\newblock
  \url{https://databricks.com/blog/2014/07/23/scalable-collaborative-filtering-with-spark-mllib.html},
  2014.
\newblock [Online; accessed 15-Aug-2015].

\bibitem{ccd++-icdm12}
H.-F. Yu, C.-J. Hsieh, S.~Si, and I.~S. Dhillon.
\newblock Scalable coordinate descent approaches to parallel matrix
  factorization for recommender systems.
\newblock In {\em ICDM}, pages 765--774, 2012.

\bibitem{nomad14}
H.~Yun, H.-F. Yu, C.-J. Hsieh, S.~Vishwanathan, and I.~S. Dhillon.
\newblock {NOMAD}: Non-locking, stochastic multi-machine algorithm for
  asynchronous and decentralized matrix completion.
\newblock In {\em VLDB}, pages 975--986, 2014.

\bibitem{Zastrau:2012:SGD}
D.~Zastrau and S.~Edelkamp.
\newblock {Stochastic gradient descent with GPGPU}.
\newblock In {\em KI 2012: Advances in Artificial Intelligence}, pages
  193--204. Springer, 2012.

\bibitem{netflix08}
Y.~Zhou, D.~M. Wilkinson, R.~Schreiber, and R.~Pan.
\newblock Large-scale parallel collaborative filtering for the netflix prize.
\newblock In {\em AAIM}, pages 337--348, 2008.

\bibitem{libmf-13}
Y.~Zhuang, W.~Chin, Y.~Juan, and C.~Lin.
\newblock A fast parallel {SGD} for matrix factorization in shared memory
  systems.
\newblock In {\em RecSys}, pages 249--256, 2013.

\end{thebibliography}
